\documentclass[twocolumn,showpacs,aps,prl]{revtex4-2}

\usepackage{amsmath,amsfonts,amssymb}
\usepackage{wrapfig}
\usepackage{graphicx}
\usepackage{bbm}
\usepackage{color}

\begin{document}

\title{Scalable Determination of Multipartite Entanglement in Quantum Networks}\date{\today}
\author{Wei-Ting Kao$^{1,2,\dag}$}
\author{Chien-Ying Huang$^{3,4,\dag}$}
\author{Tung-Ju Tsai$^{1,2}$}
\author{Shih-Hsuan Chen$^{1,2}$}
\author{Sheng-Yan Sun$^{1,2}$}
\author{Yu-Cheng Li$^{1,2}$}
\author{Teh-Lu Liao$^{1}$}
\author{Chih-Sung Chuu$^{5,6}$}
\author{He Lu$^{7}$}
\author{Che-Ming Li$^{1,2}$}
\email{cmli@mail.ncku.edu.tw}

\affiliation{$^1$Department of Engineering Science, National Cheng Kung University, Tainan 701, Taiwan}
\affiliation{$^2$Center for Quantum Frontiers of Research and Technology, National Cheng Kung University, Tainan 70101, Taiwan}
\affiliation{$^3$Department of Electrical Engineering, California Institute of Technology, Pasadena, CA 91125, USA}
\affiliation{$^4$Institute for Quantum Information and Matter, California Institute of Technology, Pasadena, CA 91125, USA}
\affiliation{$^5$Department of Physics, National Tsing Hua University, Hsinchu 30013, Taiwan}
\affiliation{$^6$Center for Quantum Science and Technology, Hsinchu 30013, Taiwan}
\affiliation{$^7$School of Physics, Shandong University, Jinan 250100, China}

\affiliation{$^{\dag}$These authors contributed equally to this work}

\begin{abstract}
Quantum networks comprised of entangled end nodes serve stronger than the classical correlation for unparalleled quantum internet applications. However, practical quantum networking is affected by noise, which at its worst, causes end nodes to be described by pre-existing classical data. In such untrusted networks, determining quantum network fidelity and genuine multi-node entanglement becomes crucial. Here, we show that determining quantum network fidelity and genuine $N$-node entanglement in an untrusted star network requires only $N+1$ measurement settings. This method establishes a semi-trusted framework, allowing some nodes to relax their assumptions. Our network determination method is enabled by detecting genuine $N$-node Einstein-Podolsky-Rosen steerability. Experimentally, using spontaneous parametric down-conversion entanglement sources, we demonstrate the determinations of genuine 3-photon and 4-photon quantum networks and the false positives of the widely used entanglement witness, the fidelity criterion of $1/2$. Our results provide a scalable method for the determination of multipartite entanglement in realistic quantum networks.
\end{abstract}

\maketitle

\section*{Introduction}
Quantum networks \cite{Kimble08,Ritter12,Wehner18} aim to provide networking nodes with the ability to process quantum information jointly, extending beyond point-to-point communication. This capability relies on quantum entanglement between the end nodes. With entanglement, quantum networks benefit from quantum correlation, coordination, and security \cite{Wehner18,Hillery99,Chen05,Markham08,Bell14,Lu16,Komar14,Proctor18,Raussendorf01,Walther05,Broadbent09,Barz12,Chen07,Lo14,Epping17}. 
Therefore, quantum networks are expected to enable applications that fundamentally enhance classical networks and serve as the building blocks of a quantum internet \cite{Northup14,Monroe14,Reiserer15,Sipahigil16,Kalb17,Pirker17,Humphreys18,Chou18,Jing19,Yu20,Stephenson20,Awschalom21,Pompili21,Lago-Rivera21,Hermans22,vanLeent22,Pompili22,Krutyanskiy23,Knaut24,Ruskuc24}.

Inevitable errors or imperfections, however, exist in required hardware components. Moreover, qubits are fragile when subjected to environmental noise. These factors can result in discrepancies between the actual networks and their target configurations. In particular, the damages can make the end nodes classical as pre-existing classical data at worst. Typically, networking participants have limited knowledge regarding such node information, rendering both the network nodes and the networking implementations untrusted from the perspective of end-users \cite{Li15,Huang19,Lu20}.

Evaluating the created networks under untrusted conditions of end nodes and hardware components is essential for realistic quantum networking. This assessment is necessary to determine whether the entanglement backbone, which underlies each quantum network stage in developing an ultimate quantum internet \cite{Kimble08,Ritter12,Wehner18}, has been faithfully established. Quantum network fidelity, also referred to as quantum fidelity or simply fidelity, is essential for evaluating the closeness of a created network state to a target state. For example, fidelity can be used to construct an entanglement witness (EW) \cite{Toth05,Horodecki09,Guhne09,McCutcheon16,Han21} to detect genuine multi-node entanglement. However, employing EWs to determine network fidelity and genuine multi-node entanglement requires assuming trusted measurement devices that adhere to the principles of quantum mechanics, rather than blindly outputting pre-existing classical data. Therefore, EWs are not applicable for determining the untrusted networking scenario.

\begin{figure}[t]
\includegraphics[width=8.7cm]{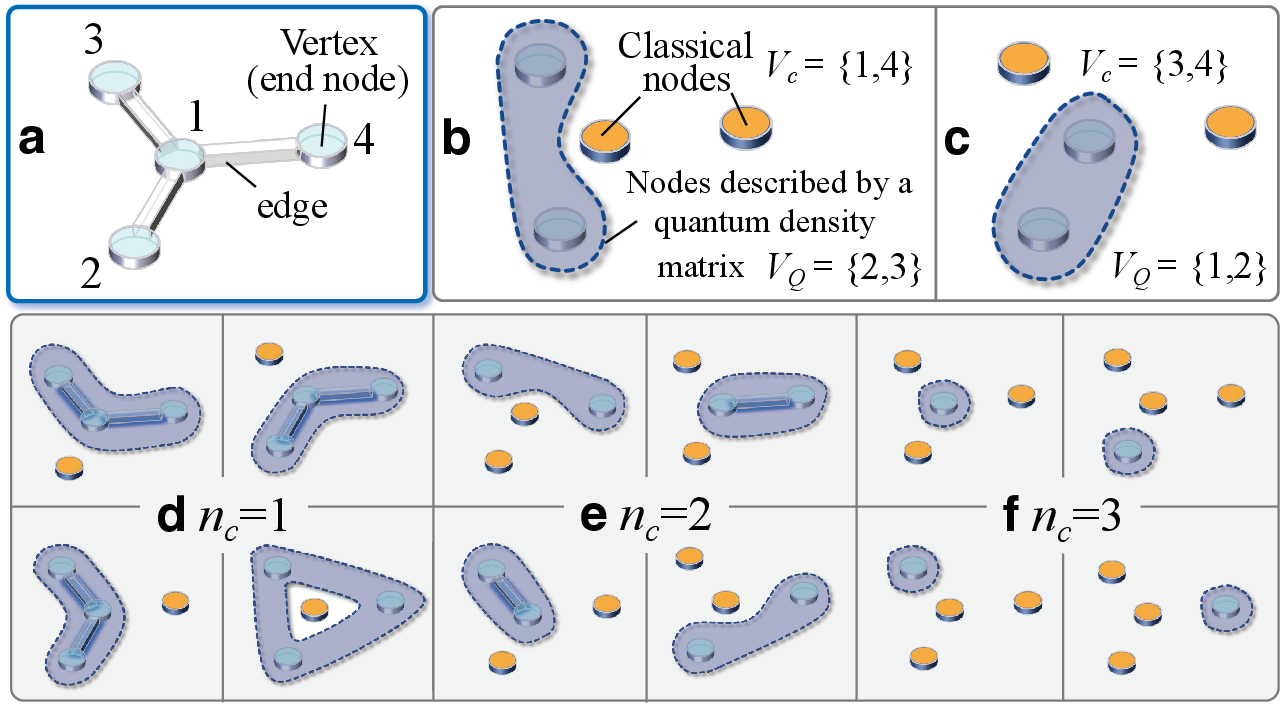}
\caption{\textbf{Distinction between quantum networks and untrusted quantum-classical hybrids.} Vertices and edges in a star topology represent an $N$-node entangled network equivalent to a GHZ state \cite{Hein04}, as illustrated in (a) for $N=4$. The former stand for end nodes (qubits), and the latter describe node-node quantum interaction. (b) A classical-quantum hybrid consists of classical nodes of pre-existing data $R_{m_{k}}$ which constitute a fixed measurement set $\textbf{v}_{k}$ and nodes described by a density matrix of quantum mechanics. See Eq.~(\ref{EPRS}). A variant can be found according to the number of classical (quantum) nodes, $n_c$ ($N-n_c$), and the combination of classical (quantum) end nodes described by the node index set $V_c$ ($V_Q$), such as (c). For $0<n_c<4$, all possible hybrids are shown in (b)-(f). We show detection of $N$-node star-topology quantum networks by ruling out all quantum-classical hybrids requires only $N+1$ measurement settings.}\label{networks}
\end{figure}

The challenge of distrust can be tackled through self-testing analysis ~\cite{Kaniewski16,Baccari20,Wu22}. However, this method only provides an estimate of the lower bound of the fidelity to the target state up to local isometries. Interestingly, Lu \textit{et al.} \cite{Lu20} demonstrated that the network fidelity could be measured in the presence of untrusted end nodes if the fidelity value exceeds the threshold for genuine multi-node Einstein-Podolsky-Rosen (EPR) steering. This approach to network determination is distinct from all other known steering identification methods  \cite{He13,Li15,Armstrong15,Cavalcanti15}.

Furthermore, as quantum networking scales up in size, the functionality of the quantum network unveils the true power of quantum application protocols, making them increasingly superior to classical networks \cite{Kimble08,Ritter12,Wehner18}. However, the challenge persists on how to conduct the necessary network determination in the presence of untrusted measurement devices with minimal experimental efforts. From a scalability standpoint, the practical steps in local measurement settings (the settings of simultaneous measurement of single qubit operators in parallel \cite{Toth05}) in the current network determination based on steering \cite{Lu20} still exhibit exponential growth with the number of end nodes.

In this work, we address the issue of distrust and scalability in network determination by detecting genuine $N$-node EPR steering in a scalable manner. This method establishes a semi-trusted framework, allowing some nodes to relax their assumptions. With only $N+1$ measurement settings, our fidelity criterion enables determining network fidelity and genuine multi-node entanglement in star network topology or Greenberger-Horne-Zeilinger (GHZ) state under untrusted conditions (Fig.~\ref{networks}). This makes our scalable network determination method useful for entanglement distribution in realistic quantum networks. The GHZ state, although not utilized in certain quantum networks such as those constructed by entangled qubit pairs \cite{Tavakoli22,Sun19,Poderini20,Suprano22,Gu23}, has played a significant and central role in various crucial quantum networking tasks. For example, the GHZ state is the resource for achieving quantum secret sharing \cite{Hillery99,Chen05,Markham08,Bell14,Lu16,Huang19} and performing one-way quantum computing \cite{Bell13}. Moreover, the GHZ state is the backbone for quantum distributed sensing \cite{Komar14,Proctor18} and quantum conference key agreement and distribution \cite{Chen07,Lo14,Epping17}. It even serves as the building block of genuine quantum networks with superposed tasks and addressing \cite{Miguel-Ramiro21}. Therefore, our formalism and the corresponding experimental demonstration help realize these assignments in quantum networks in the presence of uncharacterized network nodes.

\section*{Results}
\noindent\textbf{Scalable network determination method}\\
Suppose an $N$-node network is prepared according to an $N$-qubit graph state in a star topology \cite{Hein04}: $\left|S_{\!N}\right\rangle=( \left|0\right\rangle_1\bigotimes_{k=2}^{N} \left|+\right\rangle_k+\left|1\right\rangle_1\bigotimes_{k=2}^{N} \left|-\right\rangle_k)/\sqrt{2}$, under the first node as the center [Fig.~\ref{networks}(a)], or equivalently a GHZ state: $\left|G_{\!N}\right\rangle=(\bigotimes_{k=1}^{N}\left|0\right\rangle_k+\bigotimes_{k=1}^{N}\left|1\right\rangle_k)/\sqrt{2}$, where $\left|\pm\right\rangle_k=(\left|0\right\rangle_k\pm\left|1\right\rangle_k)/\sqrt{2}$ and $\{\left|0\right\rangle_k\equiv\left|0\right\rangle,\left|1\right\rangle_k\equiv\left|1\right\rangle\}$ is an orthonormal basis. The fidelity of a created network described by a density operator $\rho_{\text{expt}}$ and the target state $\left|\psi\right\rangle$ is defined by: $F=\text{tr}(\rho_{\text{expt}}\left|\psi\right\rangle\!\!\left\langle \psi\right|)$, for $\left|\psi\right\rangle=\left|S_{\!N}\right\rangle,\left|G_{\!N}\right\rangle$. To measure fidelity by the remote end-users, we decompose the projector into a linear combination of tensor products of $N$ local observables: $\left|\psi\right\rangle\!\!\left\langle \psi\right|=\sum_{\vec{m}}h_{\vec{m}}\bigotimes_{k=1}^{N}\hat{R}_{m_{k}}$, where $\hat{R}_{m_{k}}$ represents the $m_{k}$th observable of the $k$th qubit (end node) and $\vec{m}\equiv (m_{1},...,m_{N})$ describes the type of $N$-observable setting. The decomposition of the star-graph state or the GHZ state is not unique and decides the coefficients $h_{\vec{m}}$. To emphasize how $F$ can be measured, we rephrase the fidelity as the following fidelity function:
\begin{equation}
F(N)=\sum_{\vec{m}}h_{\vec{m}}\left\langle R_{m_{1}}...R_{m_{N}}\right\rangle,\label{F}
\end{equation}
where $R_{m_{k}}$ is the outcome of the $m_{k}$th measurement on the $k$th end node and $\left\langle R_{m_{1}}...R_{m_{N}}\right\rangle$ denote the mean value of the product of $N$ measurement outcomes. The number of local measurement settings required to measure the fidelity $F(N)$ depends on the local measurements required to determine the expectation values, $\left\langle R_{m_{1}}...R_{m_{N}}\right\rangle$.

When using the fidelity EW for entanglement detection, the $\rho_{\text{expt}}$ is detected as being genuinely $N$-qubit entangled if the measured fidelity satisfies the criterion $F(N)>\max_{\rho_{\text{B}}}\sum_{\vec{m}}h_{\vec{m}}\\\text{tr}(\rho_{\text{B}}\bigotimes_{k=1}^{N}\hat{R}_{m_{k}})=1/2$ \cite{Guhne09,Bourennane04}, which excludes the possibility of a biseparable state $\rho_{\text{B}}$. The number of local measurement settings required to obtain all the $\left\langle R_{m_{1}}...R_{m_{N}}\right\rangle$ affects the experimental efforts of determining $F$ and entanglement. When the measurements on each qubit are performed with the observables of the Pauli matrices, the total number of measurement settings is $2^{N-1}+1$ \cite{Lu20}. Whereas G\"uhne \textit{et al.} \cite{Guhne07} show that this experimental effort can be small to $N+1$ local measurement settings by introducing measurements in the $x$-$y$-plane of the Bloch sphere. This seminal determination method has been widely used in a range of multipartite entanglement experiments; see, for example, the recent determination in Refs.~\cite{Zhang22,Appel22,Wu22}.

Pre-existing classical data from classical end nodes or untrusted measurement devices can cause the resulting fidelity to be unreliable and lead to false positives of EW-based entanglement detection in quantum networks, as we will demonstrate below. We eliminate the classical possibilities and maintain the scalability of entanglement detection \cite{Guhne07} by characterizing genuine $N$-node EPR steering via the following end-node observables.

\noindent (i)~$\left|S_{\!N}\right\rangle$:
\begin{equation}
\begin{aligned}
\hat{R}_{m_{1}}&= \cos (\frac{m_{1}\pi}{N})X + \sin(\frac{m_{1}\pi}{N})Y, \  m_{1} = 1,2,..., N,\\
\hat{R}_{m_{1}} &=Z,\ \ m_{1} =N+1,\ \text{and}\\
\hat{R}_{m_{k}}&= \cos (\frac{m_{k}\pi}{N})Z - \sin(\frac{m_{k}\pi}{N})Y, \  m_{k} = 1,2,..., N,\\
\hat{R}_{m_{k}}&=X,\ \ m_{k} =N+1,\ \ \ \ \ \ \ \ \  \ {\rm for}\ k=2,3,...,N,
\end{aligned}\label{sr1}
\end{equation}
where $X$, $Y$ and $Z$ are the Pauli-$x$, Pauli-$y$, and Pauli-$z$ matrices respectively, and $X$ and $Y$ are defined via their relations to $Z\equiv\left|0\right\rangle\!\!\left\langle0\right|-\left|1\right\rangle\!\!\left\langle1\right|$.

\noindent (ii)~$\left|G_{\!N}\right\rangle$:
\begin{equation}
\begin{aligned}
\hat{R}_{m_{k}}&\!=\! \cos (\frac{m_{k}\pi}{N})X + \sin(\frac{m_{k}\pi}{N})Y,\ m_{k} = 1,2,..., N,\label{gr1}\\
\hat{R}_{m_{k}}&=Z,\ \ m_{k} =N+1, \ \ \ \ \ \ \ \ {\rm for}\ k=1,2,...,N.
\end{aligned}
\end{equation}\\
Equations~(\ref{sr1}) and (\ref{gr1}) and the identity operator $\hat{R}_{0}=I$ constitute the target state~$\left|\psi\right\rangle$, requiring only $N+1$ observables for each end node. Experimentally, using $N+1$ measurement settings is sufficient to perform the fidelity measurements~(\ref{F}) \cite{Guhne07}. See Supplementary Note 1 for concrete examples of decomposing the target GHZ states with $N+1$ observables for fidelity measurements by $N+1$ local measurement settings. Note that Eqs.~(\ref{sr1}) and (\ref{gr1}) are equivalent up to $N-1$ Hadamard transformations. This can be seen from their structures of state vectors $\left|S_{\!N}\right\rangle$ and $\left|G_{\!N}\right\rangle$. 

The $\rho_{\text{expt}}$ possesses genuine $N$-node EPR steerability close to $\left|\psi\right\rangle$ if the measured fidelity is larger than the maximum fidelity a network's measurement results mixed with the pre-existing classical data can achieve, namely
\begin{equation}
F(N)>\mathcal{F}_{c}\equiv\max_{V_c}\sum_{\vec{m}}h_{\vec{m}}\left\langle\bigotimes_{k\in V_{Q}}\hat{R}_{m_{k}}\right\rangle\prod_{k\in V_{c}}\!\!R_{m_{k}},\label{EPRS}
\end{equation}
where $V_c$ denotes the index set of classical end nodes, and $V_Q$ is the index set of the rest nodes described in quantum density matrix. See Figs.~\ref{networks}(b)-\ref{networks}(f) for classical-quantum hybrids consisting of classical nodes of pre-existing data $R_{m_{k}}$ and nodes described by a density matrix of quantum mechanics. As shown in Eq.~(\ref{F}), $h_{\vec{m}}$ represents the decomposition coefficients of target state $\left|S_{\!N}\right\rangle$ or $\left|G_{\!N}\right\rangle$ with observables~(\ref{sr1}) or (\ref{gr1}), respectively. $\hat{R}_{m_{k}}$ is the $m_{k}$th local observable of the $k$th end node belonging to $V_Q$ and following quantum mechanics. A classical end node, say the $k$th node, has physical properties independent of observation and can be specified by a fixed set of measurement outcomes, $\textbf{v}_{k}\equiv\{R_{m_{k}}|m_{k}=1,2,...,N+1\}$ where $R_{m_{k}}\in\{+1,-1\}$ corresponding to the observables~(\ref{sr1}) or (\ref{gr1}) \cite{Brunner14}. Here, the pre-existing classical data $R_{m_{k}}$ is a characteristic of a classical system. The value of the pre-existing classical data $+1$ or $-1$ represents the state of a physical property and exists before the property's measurement, independent of the observation. When considering several physical properties corresponding to observables in quantum mechanics, their classical pre-existing data constitute the pre-existing and fixed set $\textbf{v}_{k}$ that can describe the classical system. Such a classical state can be considered a pre-existing recipe fully specifying how the classical system behaves. The feature of pre-existing measurement outcomes and independent observations is also called the assumption of realism \cite{Mermin93,Guhne05,Brunner14}. Therefore, we use the realism assumption, i.e., the pre-existing data, to define and describe a node classical in noisy networks. Supplementary Note 2 provides the details of pre-existing classical data in classical nodes. The fixed set $\textbf{v}_{k}$ consisting of the pre-existing classical data $R_{m_{k}}$ under the assumption of realism has been used to calculate the fidelity upper bound, $\mathcal{F}_{c}$, for classical-quantum hybrids in Eq.~(\ref{EPRS}). As will be introduced below, we will see how the $\textbf{v}_{k}$ is used to discuss and determine $\mathcal{F}_{c}$.

Explicitly, the maximization for $\mathcal{F}_{c}$ in Eq.~(\ref{EPRS}) is over all the parameters related to $V_c$, including the classical end node number, $n_c$, for $0<n_c<N$, the combinations of these classical nodes, their pre-existing classical data of $\textbf{v}_{k}$, and the states of the quantum end nodes in $V_Q$. Therefore, we have 
\begin{equation}
\begin{aligned}
\mathcal{F}_{c}=\max_{n_c,\{\textbf{v}_{k}|k\in V_{c}\}}E[\sum_{\vec{m}}\!h_{\vec{m}}\bigotimes_{k\in V_{Q}}\hat{R}_{m_{k}} \prod_{k\in V_{c}}R_{m_{k}}], 
\end{aligned}\label{fc}
\end{equation}
where $E[\cdot]$ denotes the largest eigenvalue of the operator. After a careful derivation and calculation, we arrive at the following results:
\begin{equation}
\begin{aligned}
\mathcal{F}_{c}(N)&=\frac{1+\sqrt{1+4\frac{\csc^2(\frac{\pi}{2N})}{N^2}}}{4}  \ \ {\rm odd}\ N,\label{c}\\
\mathcal{F}_{c}(N)&=\frac{1+\sqrt{3}}{4}\simeq0.683 \ \ {\rm even}\ N.
\end{aligned}
\end{equation}
$\mathcal{F}_{c}$ with odd $N$ is monotonically decreasing: $\mathcal{F}_c(3)\simeq 0.667$, $\mathcal{F}_{c}(5) \simeq 0.6589$, and $\lim_{N\to \infty} \mathcal{F}_{c}(N) \simeq 0.6547$. Note that the equivalence between Eq.~(\ref{sr1}) and Eq.~(\ref{gr1}) leads to the same upper bounds $\mathcal{F}_{c}$ for the target states $\left|S_{\!N}\right\rangle$ and $\left|G_{\!N}\right\rangle$. To obtain Eq.~(\ref{c}) from Eq.~(\ref{fc}), we first evaluate the maximum fidelity of $\left|{G_{N}}\right\rangle$ and an $N$-node network with $n_{c}$ classical nodes: $\mathcal{F}_{n_{c}}(N)=\max_{\{\textbf{v}_{k}|k\in V_{c},|V_{c}|=n_c\}}F(N)$. Then we consider all cases where the classical nodes exist in the created network and define the maximum value of the calculated fidelities $\mathcal{F}_{n_{c}}(N)$ as $\mathcal{F}_{c}(N)$. See Supplementary Note 3 for the detailed derivation procedure of these two steps. In Supplementary Note 4, the noise tolerance of the steering criterion (\ref{EPRS}) is discussed as well.

\begin{figure}[t]
\includegraphics[width=8.7cm]{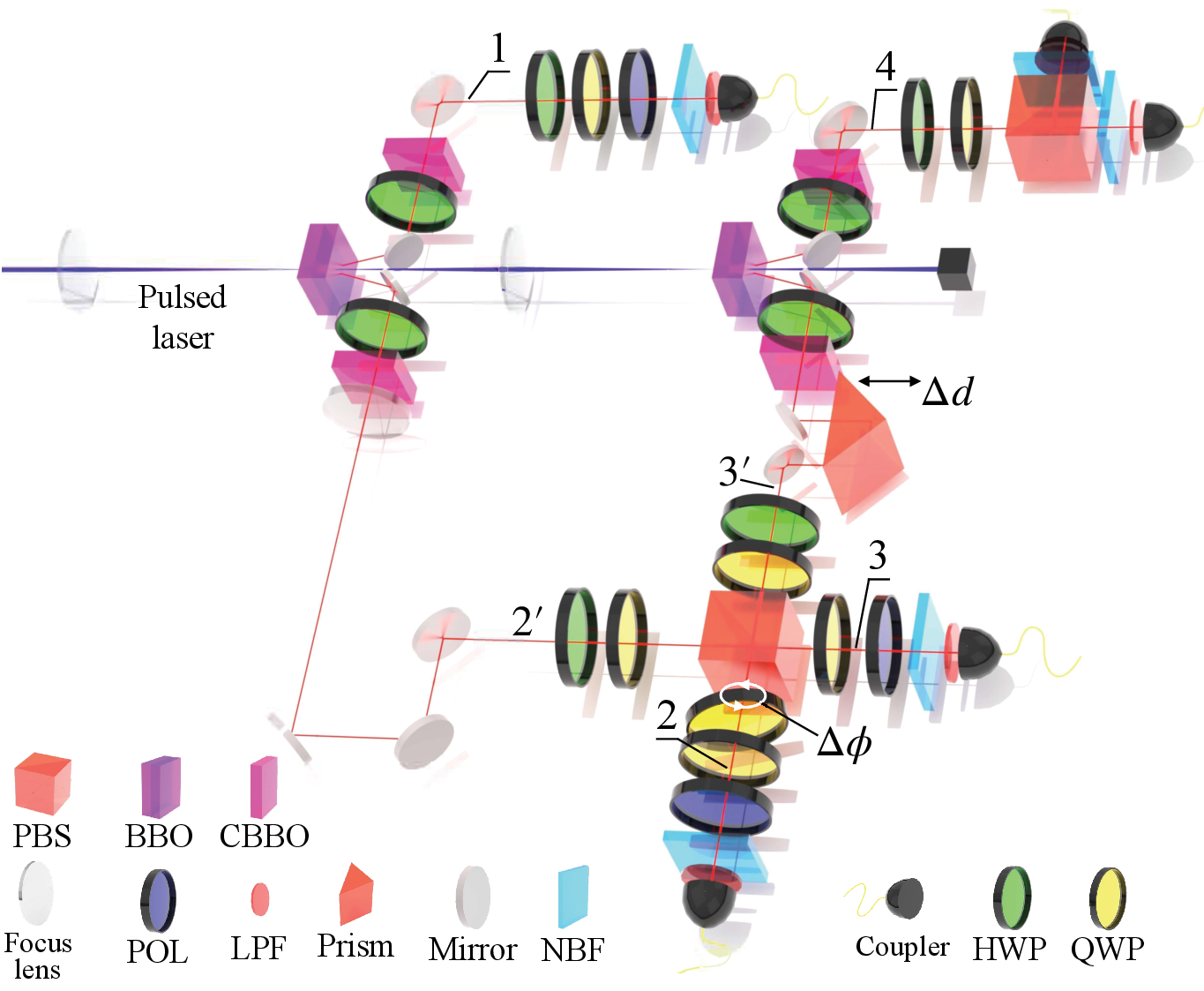}
\caption{\textbf{Experimental setup for scalable determination of multi-photon entanglement in quantum networks.} A SPDC photon source consists of a BBO crystal and a compensator [a CBBO and a half wave plate (HWP)] pumped with a pulsed laser beam. To generate multi-photon entanglement, two SPDC photon sources were created, and photons in modes $2'$ and $3'$ interfere with a PBS under good spatial and temporal overlap. Photons were selected by the narrow-band filter (NBF) and long-pass filter (LPF), collected by fiber couplers and detected by silicon avalanche single-photon detectors (not shown).}\label{setup}
\end{figure}

\begin{figure}[t]
\includegraphics[width=8.7cm]{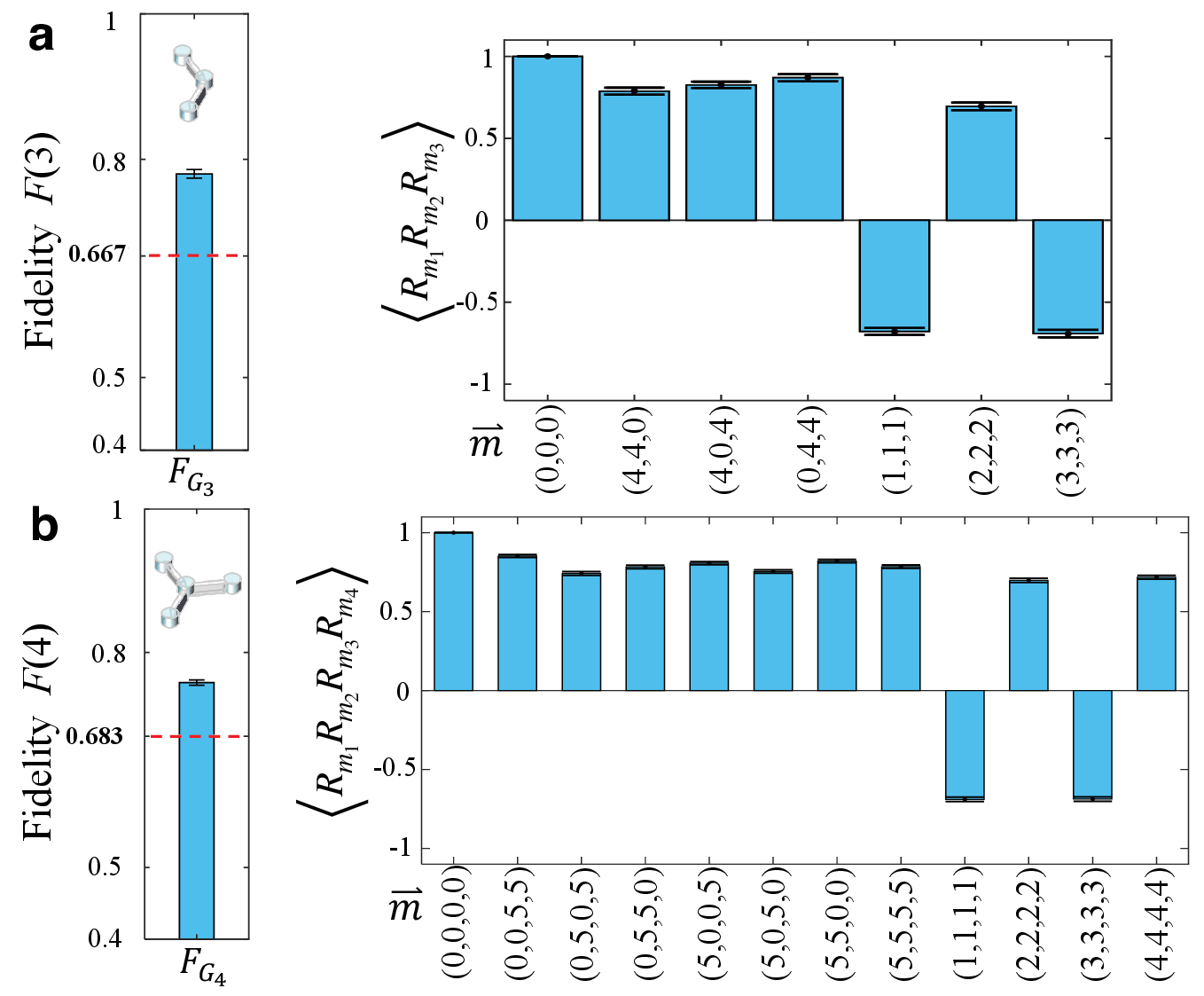}
\caption{\textbf{Experimental genuine $N$-photon EPR steerability.} We observed the experimental fidelities of (a)~$F_{G_3}(3)=(78.00\pm0.60)\%$, and (b)~$F_{G_4}(4)=(75.79\pm0.37)\%$, where the error bars represent the standard deviation of the measured photon counts based on Poissonian statistics. They reveal that the created photons possess genuine $N$-node EPR steerability under scalable $N+1$ measurement settings. Then the entanglement creation and fidelity measurements can be trusted without calibration or knowledge of all the running experimental apparatus.}\label{steering}
\end{figure}

\begin{figure}[t]
\includegraphics[width=8.7cm]{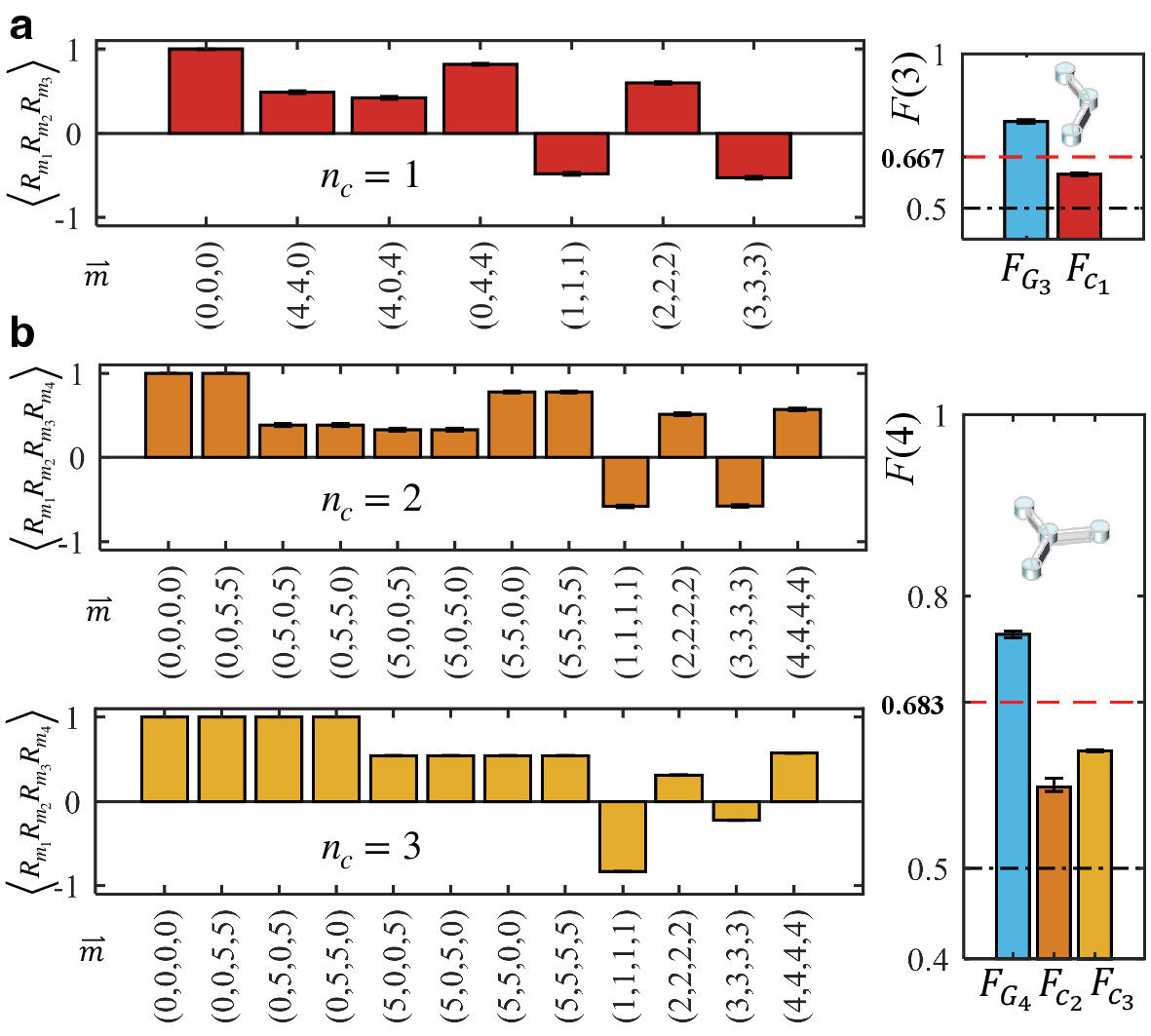}
\caption{\textbf{False positives of EW-based entanglement detection.} (a) For $N=3$, the experimental network mixed with one classical node ($n_c=1$) has the fidelity, $F_{c_1}(3)=(61.05\pm0.44)\%$, larger than than the upper bound of $1/2$ ruled out by EW-based entanglement detection. This quantum-classical hybrid was created according to the one which can achieve $\mathcal{F}_{c}(3)\simeq 0.667$ [Eq.~(\ref{c})]. (b) For $N=4$, the fidelities of the two demonstrated quantum-classical hybrid networks: $F_{c_2}(4)=(59.21\pm0.72)\%$ and $F_{c_3}(4)=(62.99\pm0.12)\%$, for $n_c=2$ and $n_c=3$, respectively, are larger than $1/2$ and smaller than $\mathcal{F}_{c}(4)\simeq 0.683$ [Eq.~(\ref{c})], where the network with three classical nodes has better ability of mimicking $\left|G_{4}\right\rangle$. These two hybrids were created aiming for the best simulations with two and three classical nodes, respectively. The error bars represent the measured photon counts' standard deviation based on Poissonian statistics.}\label{cnodes}
\end{figure}

The criterion for genuine multi-node EPR steering~(\ref{EPRS}) is stricter than the criterion for genuine multipartite entanglement. That is, the classical upper bounds (\ref{c}) imply $\mathcal{F}_{c}(N)>0.5$, which results in false positives using EW-based entanglement detection in untrusted networks \cite{McCutcheon16,Han21}. A false positive entanglement detection means a result of an EW-based entanglement test that appears to show genuine multipartite entanglement exists or is present when this is not correct. Determining network fidelity and genuine multi-node entanglement using EWs requires the assumption of trusted measurement devices that follow quantum mechanics, not outputting pre-existing classical data. Therefore, a false positive entanglement detection may occur when untrusted nodes exist in networks. As found and presented in our work, networks in the presence of classical nodes enable passing the EW-based entanglement test, which causes the false positive of genuine multipartite entanglement. As the criterion~(\ref{EPRS}) detects the steerability of $\rho_{\text{expt}}$, the possibilities of the pre-existing classical measurement outcomes from classical end nodes or untrusted measurement devices are ruled out. Therefore, the quantum network fidelity is reliably determined under untrusted measurements, and genuine multi-node entanglement is also detected.

Furthermore, our network determination method only requires $N+1$ measurement settings, which is notably different from the existing method that requires $2^{N-1}+1$ measurement settings \cite{Lu20}. For both works, the underlying concept and method to derive the maximum fidelity of networks with classical nodes depend on two items: (1) the observables used for the state decomposition of the target state, and (2) the calculation of the maximum fidelity for classical-quantum hybrids consisting of classical nodes of pre-existing data $R_{m_{k}}$ which constitute a fixed measurement set $\textbf{v}_{k}$ and nodes described by a density matrix of quantum mechanics. Lu \textit{et al.} \cite{Lu20} used Pauli matrices as observables, forming an orthonormal set resembling typical quantum state tomography. When employing orthogonal observables, the maximum fidelity of quantum-classical hybrid networks decreases monotonically with increasing classical nodes, providing a scale for network imperfections. Our method uses observables lacking this orthogonality. Explicitly, the maximum fidelity remains and does not change with the classical node number under a given odd total node number. Maximum fidelity changes with the classical node number for a given even total node number, but the fidelity value does not decrease monotonically. Moreover, the fidelity value maximized over all classical node numbers does not change with the even total node number, whereas this maximum fidelity monotonically decreases with the odd total node number. See Supplementary Table 1 for concrete examples. Without orthonormal observables, maximum fidelities do not strictly decrease with classical node count. However, the trade-off for having a scale to indicate the degree of network imperfection is the exponential increase in required measurement settings with the number of network nodes. In contrast, our method only requires a linearly increasing number of measurement settings. For a more detailed comparison between both works, see Supplementary Note 5.\\

\noindent\textbf{Experimental demonstration of genuine $N$-node EPR steering}\\
\noindent Experimentally, without loss of generality, we demonstrate the determination of genuine $3$-node and $4$-node quantum networks with EPR steering in GHZ states for the equivalence between $ \left|S_{\!N}\right\rangle$ and $ \left|G_{\!N}\right\rangle$. We created two spontaneous parametric down-conversion (SPDC) sources of polarization-entangled photon pairs and an interferometer \cite{Pan01,Pan12} to generate a multi-photon entanglement source. See Methods for more experimental details. A schematic diagram of the experimental setup is shown in Fig.~\ref{setup}. We first implemented the determination of the steerability in the created $3$-photon and $4$-photon states, $\rho_{\text{expt},G_3}$ and $\rho_{\text{expt},G_4}$, respectively. With state decomposition according to (\ref{gr1}), we performed required measurements on our created states
\begin{widetext}
\begin{eqnarray}
&&F(3)\!\!=\!\frac{1}{8}(1\!+\!\!\sum_{\pi}\!\left\langle R_{0}R_{4}R_{4}\right\rangle)\!+\!\!\frac{1}{6}\!\sum_{k=1}^{3}(-1)^{k}\!\!\left\langle R_{k}R_{k}R_{k}\right\rangle\!,\label{f3}\\
&&F(4)\!=\!\frac{1}{16}(1\!+\left\langle R_{5}R_{5}R_{5}R_{5}\right\rangle+\!\sum_{\pi}\!\left\langle R_{0}R_{0}R_{5}R_{5}\right\rangle)+\frac{1}{8}\!\sum_{k=1}^{4}(-1)^{k}\!\left\langle R_{k}R_{k}R_{k}R_{k}\right\rangle\label{f4},
\end{eqnarray}
\end{widetext}
where $\sum_{\pi}$ denotes the sum over all the permutations of the nodes. Measuring $F(3)$ and $F(4)$ requires only $4$ and $5$ measurement settings, respectively. (See Supplementary Note 1.) The experimental results are summarized in Figs.~\ref{steering}(a),~\ref{steering}(b). We generated $\rho_{\text{expt},G_3}$ and $\rho_{\text{expt},G_4}$ with $F_{G_3}(3)=(78.00\pm0.60)\%>\mathcal{F}_{c}(3)\simeq 0.667$, and $F_{G_4}(4)=(75.79\pm0.37)\%>\mathcal{F}_{c}(4)\simeq 0.683$, respectively. Our demonstrations with only $N+1$ measurement settings thus determine genuine three-photon and four-photon steerability according to the criteria (\ref{EPRS}). To demonstrate that the previous method~\cite{Lu20} requires more local measurement settings, we also obtained similar fidelities to satisfy the trustiness tests with $2^{N-1}+1$ Pauli observables. We obtained $F_{G_3}(3)=(77.93\pm0.62)$ and $F_{G_4}(4)=(75.84\pm0.31)$ using 5 and 9 local measurement settings, respectively. See Supplementary Note 8 for details of Pauli observables measurement. This shows the improved scalability of our network determination method.\\

\noindent\textbf{False positives of EW-based entanglement detection}\\
\noindent Ruling out the classical-quantum hybrid networks (Fig.~\ref{networks}) with the fidelity criterion~(\ref{EPRS}) represents genuine $N$-node entanglement in quantum networks where any classical means is impossible. One concrete illustration is to rule out the best fidelity mimicry using the pre-existing classical data that can enable false positives using EW-based entanglement detection. To experimentally demonstrate the false positives of EW-based entanglement detection in $N=3$ and $n_c=1$ network, we created a 2-photon state mixed with pre-existing classical data from 1 classical node. This quantum-classical hybrid was created according to the one which can achieve the best fidelity mimicry $\mathcal{F}_{c}(3)\simeq 0.667$ [Eq.~(\ref{c})]. See Methods for more details about the preparation and measurement of quantum-classical hybrids. As shown in Fig.~\ref{cnodes}(a), we observed the fidelity of the quantum-classical hybrid $F_{c_1}(3)=(61.05\pm0.44)\%$, which is greatly larger than the widely used EW fidelity criterion of $1/2$ for the detection of genuine $N$-node entanglement \cite{Horodecki09,Guhne09,McCutcheon16,Han21}. Whereas the created 2-photon state mixed with 1 classical node shows the false positive of detecting genuine $3$-node entanglement using EW-based entanglement detection. Our fidelity criterion through $\mathcal{F}_{c}(3)$ rules out the mimicry of genuine $3$-node entanglement with a quantum-classical hybrid.

Similarly, we created $N=4$ quantum-classical hybrid networks using 2-photon state and 1-photon state for $n_c=2$ and $n_c=3$, respectively. We observed the fidelities $F_{c_2}(4)=(59.21\pm0.72)\%$ and $F_{c_3}(4)=(62.99\pm0.12)\%$, for $n_c=2$ and $n_c=3$, respectively, which are larger than $1/2$ and smaller than $\mathcal{F}_{c}(4)\simeq 0.683$, as shown in Figs.~\ref{cnodes}(b),(c). Our demonstrations validated the false positives of EW-based entanglement detection and the practicality of our determination method in quantum networks. It is worth emphasizing that environmental noise or experimental imperfections are generally unknown in realistic networking, and more than one such extreme case, including their mixtures, must be excluded. Therefore satisfying the fidelity criterion~(\ref{EPRS}) becomes significant to rule out all possibilities of false positives.

\section*{Discussion}

The presented results provide an efficient method for the determination the quantum network fidelity and genuine multi-node entanglement in the presence of untrusted nodes with the minimum experimental effort in existing protocols. They benefit from detecting genuine $N$-node EPR steering in star topology networks with only $N$+1 measurement settings in a semi-trusted framework where some nodes can relax their quantum mechanical assumptions. The proof-of-principle photonic network experiments validate that the maximum fidelity obtained from the combination of measurement results of trusted nodes and the pre-existing classical data can surpass the widely used fidelity criterion of $1/2$ \cite{Horodecki09,Guhne09,McCutcheon16,Han21}. This clearly shows the false positives of EW-based entanglement detection in the presence of untrusted measurement devices. Our results, therefore, offer a scalable manner to faithfully determine multipartite entanglement in quantum networks and insights into evaluating the entanglement backbone in realistic quantum networks \cite{Northup14,Monroe14,Reiserer15,Sipahigil16,Kalb17,Pirker17,Humphreys18,Chou18,Jing19,Yu20,Stephenson20,Awschalom21,Pompili21,Lago-Rivera21,Hermans22,vanLeent22,Pompili22,Krutyanskiy23,Knaut24,Ruskuc24}. We expect our method could be extended to other types of network topology for generic quantum internet with further studies.

\section*{Methods}
\noindent\textbf{Multi-photon entanglement source}\\
The two SPDC sources were pumped with a pulsed laser beam (center wavelength 390 nm, pulse duration 144 fs, repetition rate 76 MHz, and average power 1.0 W). In each source, a $\beta$-barium-borate (BBO) crystal ($2$ mm) was pumped to generate pairs of polarization-entangled photons by type-II non-collinear SPDC, including the compensation for the walk-off effect by an additional BBO crystal (CBBO, 1 mm). See Fig.~\ref{setup} for the experimental setup. Supplementary Note 6 details the laser system and photon source. The prepared two photon pairs in modes 1-$2'$ and $3'$-4 have an average creation rate of $\sim99\times10^3$ pairs per second and an average fidelity of $(93.61\pm0.09)\%$ close to $\left|\Phi^+\right\rangle=(\left|HH\right\rangle+\left|VV\right\rangle)/\sqrt{2}$, where $\left|H\right\rangle\equiv\left|0\right\rangle$ ($\left|V\right\rangle\equiv\left|1\right\rangle$) denotes the horizontal (vertical) photon polarization state. Then, photons in modes $2'$ and $3'$ were made to temporally and spatially overlap~\cite{HOM87} at a polarizing beam splitter (PBS) by fine-tuning the delay $\Delta d$. The observed visibility of the Hong-Ou-Mandel (HOM) type interference fringes is better than $76\%$. Supplementary Note 7 provides more details of HOM experiments. By properly tuning the angle $\Delta\phi$ of the quarter wave plate (QWP) around its vertically-oriented fast axis and after post-selected four-fold coincidence detection, a four-photon experimental state $\rho_{\text{expt},G_4}$ in mode 1-2-3-4 was created at a rate of $\sim 76$ events per second. A three-photon state $\rho_{\text{expt},G_3}$ in mode 2-3-4 was generated in the same manner as the photon in mode 1 was measured as a trigger.\\  

\noindent\textbf{Preparation and measurement of quantum-classical hybrids}\\ 
To experimentally create quantum-classical hybrids for false positives of EW-based entanglement detection, we prepared the eigenstates under specific conditions of $V_c$ and $\{\textbf{v}_{k}| k\in V_{c}\}$ for the fidelity operator $\hat{F}_{n_c}(N)$ (see Eq.~(16) in Supplementary Note 3), which correspond to the best fidelity mimicry of the target state $\left|G_{\!N}\right\rangle$. For $N=3$ and $n_c=1$ network with the state decomposition of $\left|G_{\!3}\right\rangle$ (see Eq.~(4) in Supplementary Note 1), we have the fidelity operator $\hat{F}_{n_c}(N)$ for a two-photon state in modes 3 and 4, under the classical node in mode 2 of $\textbf{v}_{2}=\{+1,+1,+1,+1\}$ (see Eq.~(9) in Supplementary Note 2), 
\begin{widetext}
\begin{eqnarray}
\hat{F}_{1}(3)=\frac{1}{8}( I \otimes I+  I \otimes Z +  Z \otimes Z+  Z \otimes I) + \frac{1}{6}(-\hat{R}_{1} \otimes \hat{R}_{1}+ \hat{R}_{2} \otimes \hat{R}_{2}- \hat{R}_{3} \otimes\hat{R}_{3}).\label{f1}
\end{eqnarray} 
\end{widetext}
We then created two-photon state in modes $3$ and $4$ according to the eigenvector with the maximum eigenvalue $\mathcal{F}_{1}(3)$ of Eq.~(\ref{f1}). Explicitly, such an eigenvector is of the form: 
\begin{equation}
\left|e_{1}(3)\right\rangle=\frac{1}{\sqrt{|\alpha|^2+1}}(\alpha\left|HH\right\rangle+\left|VV\right\rangle),
\end{equation}
where $\alpha=-1+i\sqrt{3}$. Note that under this condition, $\mathcal{F}_{1}(3)=\mathcal{F}_{c}(3)\simeq 0.667$. Supplementary Note 3 provides the detailed derivation of $\mathcal{F}_{c}$ [Eq.~(\ref{c})] from $\hat{F}_{n_c}(N)$. Experimentally, the two-photon state was created by measuring the photon in mode $2$ of the state $\rho_{\text{expt},G3}$ via a QWP setting at $-16.85^{\circ}$ and a polarizer (POL) setting at $5.08^{\circ}$ (Fig.~\ref{setup}). With $\textbf{v}_{2}=\{+1,+1,+1,+1\}$ we observed a fidelity of $F_{c_1}=(61.05\pm0.44)\%$, which is larger than the upper bound of $1/2$ ruled out by EW-based entanglement detection .

As the same method used above, for $N=4$ and $n_c=2$, with the state decomposition of $\left|G_{\!4}\right\rangle$ and $\textbf{v}_{1}=\{+1,+1,+1,+1,+1\}$ and $\textbf{v}_{2}=\{-1,+1,+1,-1,+1\}$ (see Eqs.~(5) and (10) in Supplementary Note 2) we can write the fidelity operator $\hat{F}_{n_c}(N)$ for a two-photon state in modes 3 and 4,
\begin{widetext}
\begin{eqnarray}
\hat{F}_{2}(4)&=&\frac{1}{16}(I\otimes I+ I\otimes I + I\otimes Z+ I\otimes Z+ Z\otimes I+ Z\otimes Z+ Z\otimes I+ Z\otimes Z)\nonumber\\
&&\ \ \ + \frac{1}{8}(\hat{R}_{1}\otimes\hat{R}_{1}+\hat{R}_{2}\otimes\hat{R}_{2}-\hat{R}_{3}\otimes\hat{R}_{3}-\hat{R}_{4}\otimes\hat{R}_{4}).
\end{eqnarray}
\end{widetext}
We then generated a two-photon state in modes $3$ and $4$ according to the eigenvector of $\hat{F}_{2}(4)$ with the maximum eigenvalue, $\mathcal{F}_{2}(4)=\mathcal{F}_{c}(4)\simeq0.683$:
\begin{equation}
\left|e_{2}(4)\right\rangle=\frac{1}{\sqrt{|\beta|^2+1}}(\beta\left|HH\right\rangle+\left|VV\right\rangle),
\end{equation}
where $\beta=(1+\sqrt{3})e^{i\pi/4}/\sqrt{2}$. The two-photon state was created by measuring the photon in mode $2$ of the state $\rho_{\text{expt},G3}$ by setting QWP and POL at $25.18^{\circ}$ and $34.02^{\circ}$, respectively. Combined with $\textbf{v}_{1}=\{+1,+1,+1,+1,+1\}$ and $\textbf{v}_{2}=\{-1,+1,+1,-1,+1\}$, the experimental value of the fidelity is $F_{c_2}=(59.21\pm0.72)\%$, which shows the false positive of detecting genuine $3$-node entanglement using EW-based entanglement detection.

Furthermore, we demonstrated the false positives of EW-based entanglement detection by creating a state close to a target of one photon in mode $1$:
\begin{equation}
\left|e_{3}(4)\right\rangle=\frac{1}{\sqrt{|\gamma|^2+1}}(\gamma\left|H\right\rangle+\left|V\right\rangle),
\end{equation}
where $\gamma=[2+\sqrt{2(4+\sqrt{2})}]/(1+\sqrt{2}+i)$, and three classical nodes with the states $\textbf{v}_{2}=\{+1,+1,+1,+1,+1\}$, $\textbf{v}_{3}=\{+1,+1,+1,+1,+1\}$ and $\textbf{v}_{4}=\{-1,+1,+1,+1,-1\}$ (see Eq.~(10) in Supplementary Note 2), which provides $\mathcal{F}_{3}(4)\simeq0.661$. Such a state was prepared by setting QWP and POL at $0^{\circ}$ and $27.38^{\circ}$, respectively, and then measuring the photon in mode $2'$ of the experimental two-photon state, $\rho_{\text{expt},\Phi^+}$. With the created state, we observed $F_{c_3}=(62.99\pm0.12)\%$, which demonstrated the false positive of EW-based entanglement detection.

We thank S.-L. Chen, Y.-N. Chen, Y.-A. Chen, and J.-W. Pan for helpful comments and discussions. C.-Y. Huang was partially supported by the MOE Taiwan-Caltech Fellowship. This work was partially supported by the National Science and Technology Council, Taiwan, under Grant Numbers MOST 107-2628-M-006-001-MY4, MOST 111-2112-M-006-033, MOST 111-2119-M-007-007, MOST 111-2123-M-006-001, and NSTC 112-2112-M-006-029.

\bibliography{SDMEQN}

\end{document}

\title{Supplementary Information for\\ Scalable Determination of Multipartite Entanglement in Quantum Networks}

\author{Wei-Ting Kao$^{1,2,\dag}$}
\author{Chien-Ying Huang$^{3,4,\dag}$}
\author{Tung-Ju Tsai$^{1,2}$}
\author{Shih-Hsuan Chen$^{1,2}$}
\author{Sheng-Yan Sun$^{1,2}$}
\author{Yu-Cheng Li$^{1,2}$}
\author{Teh-Lu Liao$^{1}$}
\author{Chih-Sung Chuu$^{5,6}$}
\author{He Lu$^{7}$}
\author{Che-Ming Li$^{1,2}$}
\email{cmli@mail.ncku.edu.tw}

\affiliation{$^1$Department of Engineering Science, National Cheng Kung University, Tainan 701, Taiwan}
\affiliation{$^2$Center for Quantum Frontiers of Research and Technology, National Cheng Kung University, Tainan 70101, Taiwan}
\affiliation{$^3$Department of Electrical Engineering, California Institute of Technology, Pasadena, CA 91125, USA}
\affiliation{$^4$Institute for Quantum Information and Matter, California Institute of Technology, Pasadena, CA 91125, USA}
\affiliation{$^5$Department of Physics, National Tsing Hua University, Hsinchu 30013, Taiwan}
\affiliation{$^6$Center for Quantum Science and Technology, Hsinchu 30013, Taiwan}
\affiliation{$^7$School of Physics, Shandong University, Jinan 250100, China}

\affiliation{$^{\dag}$These authors contributed equally to this work}

\maketitle

\section*{SUPPLEMENTARY NOTE 1: Target state decomposition}

To measure fidelity by the remote end-users, the target state projector is decomposed into a linear combination of tensor products of $N$ local observables. Here we illustrate the decomposition of the target Greenberger-Horne-Zeilinger (GHZ) state projector $\left|{G_{N}}\right\rangle\!\!\left\langle {G_{N}}\right|$ by using $N+1$ observables \cite{Guhne07}:

\begin{eqnarray}
\left|{G_{N}}\right\rangle\!\!\left\langle {G_{N}}\right|=\frac{1}{2^N}\sum_{\vec{m}|m_k\in\{0,N+1\}}c_{\vec{m}}\bigotimes^N_{k=1}\hat{R}_{m_{k}} +\frac{1}{2N}\sum_{k=1}^{N}(-1)^k\hat{R}_{k}^{\otimes N},\label{dgn}
\end{eqnarray}
where $\vec{m}\equiv(m_{1}, ..., m_{N})$, $c_{\vec{m}}= \text{tr}(\left|{G_{N}}\right\rangle\!\!\left\langle {G_{N}}\right| \bigotimes^N_{k=1}\hat{R}_{m_{k}})\in\{0,1\}$, and
\begin{equation}
\begin{aligned}
\hat{R}_{m_{k}}&= \cos (\frac{m_{k}\pi}{N})X + \sin(\frac{m_{k}\pi}{N})Y, \ m_{k} = 1,..., N,\label{gr1}\\
\hat{R}_{m_{k}}&=Z,\ \ m_{k} =N+1.
\end{aligned}
\end{equation}
Equation~(\ref{gr1}) describes the used $N+1$ observables; see also Eq.~(3) in the main text. Here $c_{\vec{m}}/2^N$ and $(-1)^k/2N$ are the decomposition coefficients, $h_{\vec{m}}$, as shown in the fidelity function $F(N)$ [Eq.~(1) in the main text]. With the decomposition~(\ref{dgn}), the corresponding fidelity function is
\begin{equation}
F(N)=\frac{1}{2^N}\sum_{\vec{m}|m_k\in\{0,N+1\}}c_{\vec{m}}\left\langle R_{m_{1}}...R_{m_{N}}\right\rangle +\frac{1}{2N}\sum_{k=1}^{N}(-1)^k\left\langle R_{k}...R_{k}\right\rangle.\label{dgnf}
\end{equation}
When experimentally measuring $F(N)$, we need one local measurement setting to determine $\sum_{\vec{m}|m_k\in\{0,N+1\}}c_{\vec{m}}\left\langle R_{m_{1}}...R_{m_{N}}\right\rangle$ and $N$ local measurement settings to determine $\frac{1}{2N}\sum_{k=1}^{N}(-1)^k\left\langle R_{k}...R_{k}\right\rangle$. Therefore, the total number of measurement settings is $N+1$. In what follows, we provide two concrete examples to illustrate the decomposition of $\left|{G_{N}}\right\rangle\!\!\left\langle {G_{N}}\right|$ shown in Eq.~(\ref{dgn}).

\subsection*{Examples: $\left|{G_{3}}\right\rangle$ and $\left|{G_{4}}\right\rangle$}

For $\left|{G_{3}}\right\rangle$ and $\left|{G_{4}}\right\rangle$, they have the following decompositions:
\begin{eqnarray}
\begin{aligned}
\left|{G_3}\right\rangle\!\!\left\langle {G_3}\right| &= \frac{1}{8}(\hat{R}_0\otimes \hat{R}_0\otimes \hat{R}_0+ \hat{R}_0\otimes \hat{R}_4\otimes \hat{R}_4+ \hat{R}_4\otimes \hat{R}_4\otimes \hat{R}_0+ \hat{R}_4\otimes \hat{R}_0\otimes \hat{R}_4)\\
&\ +\frac{1}{6}(-\hat{R}_{1}\otimes\hat{R}_{1}\otimes\hat{R}_{1}+\hat{R}_{2}\otimes\hat{R}_{2}\otimes\hat{R}_{2}-\hat{R}_{3}\otimes\hat{R}_{3}\otimes\hat{R}_{3}),\label{gm3}
\end{aligned}
\end{eqnarray}
where $\hat{R}_0=I$, and
\begin{eqnarray}
\begin{aligned}
\left|{G_{4}}\right\rangle\!\!\left\langle {G_{4}}\right| &= \frac{1}{16}(\hat{R}_0\otimes \hat{R}_0\otimes \hat{R}_0\otimes \hat{R}_0+\hat{R}_0\otimes \hat{R}_0\otimes\hat{R}_5\otimes \hat{R}_5+ \hat{R}_0\otimes\hat{R}_5 \otimes\hat{R}_0 \otimes\hat{R}_5+ \hat{R}_0\otimes\hat{R}_5\otimes\hat{R}_5 \otimes\hat{R}_0 \\
&\ \ \ \ \ \ +\hat{R}_5\otimes\hat{R}_0 \otimes\hat{R}_5 \otimes\hat{R}_0+ \hat{R}_5\otimes\hat{R}_5\otimes\hat{R}_0 \otimes\hat{R}_0+ \hat{R}_5\otimes\hat{R}_0 \otimes\hat{R}_0\otimes\hat{R}_5
+ \hat{R}_5 \otimes\hat{R}_5 \otimes\hat{R}_5 \otimes\hat{R}_5)\\
&\ \ \ +\frac{1}{8}(-\hat{R}_{1}\otimes \hat{R}_{1} \otimes \hat{R}_{1} \otimes \hat{R}_{1} + \hat{R}_{2} \otimes\hat{R}_{2} \otimes\hat{R}_{2} \otimes \hat{R}_{2} -\hat{R}_{3} \otimes \hat{R}_{3} \otimes \hat{R}_{3} \otimes \hat{R}_{3}+ \hat{R}_{4} \otimes \hat{R}_{4} \otimes \hat{R}_{4} \otimes \hat{R}_{4}).\label{gm4}
\end{aligned}
\end{eqnarray}
They have been used to demonstrate genuine $N$-node EPR steering and false positives of EW-based entanglement detection in our experiments. Their corresponding fidelity functions are shown in Eqs.~(7) and (8), respectively, in the main text. For $F(3)$ [Eq.~(7)], determining $\sum_{\pi}\!\left\langle R_{0}R_{4}R_{4}\right\rangle$ and $\sum_{k=1}^{3}(-1)^{k}\!\!\left\langle R_{k}R_{k}R_{k}\right\rangle$ requires one measurement setting and three measurement settings, respectively. For $F(4)$ [Eq.~(8)], determining $\left\langle R_{5}R_{5}R_{5}R_{5}\right\rangle+\!\sum_{\pi}\!\left\langle R_{0}R_{0}R_{5}R_{5}\right\rangle)$ and $\sum_{k=1}^{4}(-1)^{k}\!\left\langle R_{k}R_{k}R_{k}R_{k}\right\rangle$ needs one measurement setting and four measurement settings, respectively. Therefore, measuring $F(3)$ and $F(4)$ requires $4$ and $5$ measurement settings, respectively.\\

\subsection*{Comparison: target state decomposition with the Pauli observables~\cite{Lu20}}

Compared to measuring fidelity with the Pauli observables, $2^{N-1}+1$ measurement settings \cite{Lu20} are required to perform fidelity measurements according to the decomposition of $\left|{G_{N}}\right\rangle\!\!\left\langle {G_{N}}\right|$. For $N=3$, we need $5$ different types of local measurements to measure $F(3)$ according to the following state decomposition:
\begin{eqnarray}
\begin{aligned}
\left|{G_{3}}\right\rangle\!\!\left\langle {G_{3}}\right|=&\dfrac{1}{8}({I}\otimes I\otimes I+{I}\otimes{Z}\otimes{Z}+{Z}\otimes{I}\otimes{Z}+{Z}\otimes{Z}\otimes{I}\\
&\ \ +{X}\otimes X\otimes X-{X}\otimes{Y}\otimes{Y}-{Y}\otimes{X}\otimes{Y}-{Y}\otimes{Y}\otimes{X}).
\end{aligned}\label{G3_pauli}
\end{eqnarray}
As $N=4$, 9 local measurements are necessary for measuring $F(4)$ according to the state decomposition:
\begin{eqnarray}
\begin{aligned}
\label{G4_pauli}
\left|{G_{4}}\right\rangle\!\!\left\langle {G_{4}}\right|=&\frac{1}{16}({I}\otimes I\otimes I\otimes I+{Z}\otimes Z\otimes Z\otimes Z+{I}\otimes{I}\otimes{Z}\otimes{Z}+{I}\otimes{Z}\otimes{I}\otimes{Z}\\
&\ \ \ \ +{Z}\otimes{I}\otimes{I}\otimes{Z}+{Z}\otimes{I}\otimes{Z}\otimes{I}+{I}\otimes{Z}\otimes{Z}\otimes{I}+{Z}\otimes{Z}\otimes{I}\otimes{I} \\
&\ \ \ \ +{X}\otimes X\otimes X\otimes X+{Y}\otimes Y\otimes Y\otimes Y-{Y}\otimes{Y}\otimes{X}\otimes{X}-{X}\otimes{Y}\otimes{Y}\otimes{X}\\
&\ \ \ \ -{X}\otimes{X}\otimes{Y}\otimes{Y}-{Y}\otimes{X}\otimes{Y}\otimes{X}-{Y}\otimes{X}\otimes{X}\otimes{Y}-{X}\otimes{Y}\otimes{X}\otimes{Y}).
\end{aligned}
\end{eqnarray}\\

\section*{SUPPLEMENTARY NOTE 2: Classical nodes in networks}

The variables of a classical node exist before the measurement, independent of the observation, called the assumption of realism \cite{Brunner14}. This gives the pre-existing classical data as experimental results. Such pre-existing classical data can also be from the blind output of untrusted measurement devices. In contrast, quantum nodes may not have definite physical property values.

We suppose that the created $N$-node network consisting of $n_Q$ quantum nodes and $n_c$ classical nodes, where $N = n_{Q} + n_{c}$. The index set of network nodes, $V$, can therefore be divided into a subset of quantum nodes, $V_Q$, and a subset of classical nodes, $V_c$, accordingly, with $\left| V_c \right|= n_c$ and $\left| V_Q \right|= n_Q$. See Figs.~1(b)-1(f) in the main text for concrete illustrations. If the $k$th node is classical, the measurement outcomes on the node can be specified by a pre-existing and fixed set corresponding to the $N+1$ observables~(\ref{gr1}):
\begin{eqnarray}
\textbf{v}_{k}\equiv\{ R_1,R_2,...,R_{N+1}\},\label{aai}
\end{eqnarray}
for $k\in V_{c}$. With the pre-existing classical data $R_{m_{k}}\in \{+1,-1\}$, there are $2^{N+1}$ possible kinds of pre-existing sets. Let us take $N=3$ for example, there are following $2^4=16$ pre-existing sets:
\begin{eqnarray}
&&\textbf{v}_{k,1}=\{+1,+1,+1,+1\}, \textbf{v}_{k,2}=\{-1,+1,+1,+1\}, \nonumber \\
&&\textbf{v}_{k,3}=\{+1,-1,+1,+1\}, \textbf{v}_{k,4}=\{+1,+1,-1,+1\}, \nonumber \\
&&\textbf{v}_{k,5}=\{+1,+1,+1,-1\}, \textbf{v}_{k,6}=\{-1,-1,+1,+1\}, \nonumber \\
&&\textbf{v}_{k,7}=\{-1,+1,-1,+1\}, \textbf{v}_{k,8}=\{-1,+1,+1,-1\},\nonumber  \\
&&\textbf{v}_{k,9}=\{+1,-1,-1,+1\}, \textbf{v}_{k,10}=\{+1,-1,+1,-1\}, \label{v3} \\ 
&&\textbf{v}_{k,11}=\{+1,+1,-1,-1\}, \textbf{v}_{k,12}=\{-1,-1,-1,+1\}, \nonumber \\
&&\textbf{v}_{k,13}=\{+1,-1,-1,-1\}, \textbf{v}_{k,14}=\{-1,+1,-1,-1\}, \nonumber \\
&&\textbf{v}_{k,15}=\{-1,-1,+1,-1\}, \textbf{v}_{k,16}=\{-1,-1,-1,-1\}.\nonumber 
\end{eqnarray}
We apply $\textbf{v}_{2,1}$ to Eq.~(9) in Methods for demonstrating false positives of EW-based entanglement detection in the main text.
Moreover, for $N=4$, we have $2^5=32$ possible sets of pre-existing data:
\begin{eqnarray}
&&\textbf{v}_{k,1}=\{+1,+1,+1,+1,+1\}, \textbf{v}_{k,2}=\{-1,+1,+1,-1,+1\}, \nonumber \\
&&\textbf{v}_{k,3}=\{+1,+1,+1,+1,-1\}, \textbf{v}_{k,4}=\{+1,+1,+1,-1,+1\}, \nonumber \\
&&\textbf{v}_{k,5}=\{+1,+1,-1,+1,+1\}, \textbf{v}_{k,6}=\{+1,-1,+1,+1,+1\}, \nonumber \\
&&\textbf{v}_{k,7}=\{-1,+1,+1,+1,+1\}, \textbf{v}_{k,8}=\{-1,-1,+1,+1,+1\},\nonumber  \\
&&\textbf{v}_{k,9}=\{-1,+1,-1,+1,+1\}, \textbf{v}_{k,10}=\{+1,-1,+1,+1,-1\}, \nonumber \\
&&\textbf{v}_{k,11}=\{-1,+1,+1,+1,-1\}, \textbf{v}_{k,12}=\{+1,-1,-1,+1,+1\}, \nonumber \\
&&\textbf{v}_{k,13}=\{+1,-1,+1,-1,+1\}, \textbf{v}_{k,14}=\{+1,+1,-1,-1,+1\}, \nonumber \\
&&\textbf{v}_{k,15}=\{+1,+1,-1,+1,-1\}, \textbf{v}_{k,16}=\{+1,+1,+1,-1,-1\},\nonumber \\
&&\textbf{v}_{k,17}=\{+1,+1,-1,-1,-1\}, \textbf{v}_{k,18}=\{+1,-1,+1,-1,-1\},  \label{v4} \\
&&\textbf{v}_{k,19}=\{+1,-1,-1,+1,-1\}, \textbf{v}_{k,20}=\{+1,-1,-1,-1,+1\}, \nonumber \\
&&\textbf{v}_{k,21}=\{-1,+1,+1,-1,-1\}, \textbf{v}_{k,22}=\{-1,+1,-1,+1,-1\}, \nonumber \\
&&\textbf{v}_{k,23}=\{-1,+1,-1,-1,+1\}, \textbf{v}_{k,24}=\{-1,-1,+1,+1,-1\},\nonumber  \\
&&\textbf{v}_{k,25}=\{-1,-1,+1,-1,+1\}, \textbf{v}_{k,26}=\{-1,-1,-1,+1,+1\}, \nonumber \\
&&\textbf{v}_{k,27}=\{+1,-1,-1,-1,-1\}, \textbf{v}_{k,28}=\{-1,+1,-1,-1,-1\}, \nonumber \\
&&\textbf{v}_{k,29}=\{-1,-1,+1,-1,-1\}, \textbf{v}_{k,30}=\{-1,-1,-1,+1,-1\}, \nonumber \\
&&\textbf{v}_{k,31}=\{-1,-1,-1,-1,+1\}, \textbf{v}_{k,32}=\{-1,-1,-1,-1,-1\}.\nonumber
\end{eqnarray}
We apply $\textbf{v}_{1,1}, \textbf{v}_{2,2}$ to Eq.~(11) and $\textbf{v}_{2,1}, \textbf{v}_{3,1}, \textbf{v}_{4,11}$ to simulate $\mathcal{F}_{3}(4)\simeq0.661$ in Methods for illustrating false positives of EW-based entanglement detection in the main text.\\

\section*{SUPPLEMENTARY NOTE 3: Fidelity criteria and classical upper bounds $\mathcal{F}_{c}(N)$}

The fidelity criterion~(4) proposed in the main text confirms that the measured fidelity is larger than the maximum fidelity a network's measurement results mixed with the pre-existing classical data can achieve, $\mathcal{F}_{c}(N)$ [Eq.~(6) in the main text]. Here, we will show how to derive $\mathcal{F}_{c}(N)$ for GHZ states. According to the following analysis procedure consisting of two steps in sub-sections~A and B, we describe how to determine the fidelity via pre-existing measurement data sets~(\ref{aai}) and show the maximum fidelity, $\mathcal{F}_{c}(N)$, that can be achieved by these pre-existing classical sets for the target GHZ states.\\

\subsection*{Maximum classical fidelity under given $n_c$: $\mathcal{F}_{n_{c}}(N)$}

Determining the classical upper bounds shown in the fidelity criterion~(4) requires performing the maximization task:
\begin{equation}
\mathcal{F}_{c}\equiv\max_{V_c}\sum_{\vec{m}}h_{\vec{m}}\left\langle\bigotimes_{k\in V_{Q}}\!\!\!\hat{R}_{m_{k}}\right\rangle\prod_{k\in V_{c}}\!\!R_{m_{k}}.\label{EPRS}
\end{equation}
We first consider the contribution from the parameter of the number of classical nodes, $n_c$. We evaluate the maximum fidelity of $\left|{G_{N}}\right\rangle$ and an $N$-node network with $n_{c}$ classical nodes:
\begin{equation}
\mathcal{F}_{n_{c}}(N)=\max_{\{\textbf{v}_{k}|k\in V_{c},|V_{c}|=n_c\}}F(N),\label{aal}
\end{equation}
where the maximization is for all vertex sets $V_c$ with $\left| V_c \right|= n_c$ that exist in the network and all possible measurement outcomes of the quantum nodes and classical nodes. We follow three steps given below for this maximization: \\

\noindent(Step 1) An ideal network with GHZ state \emph{before} decaying into a quantum-classical hybrid can be rephrased according to the classification of classical and quantum node after decaying as

\begin{eqnarray}
\left|{G_{N}}\right\rangle\!\!\left\langle {G_{N}}\right|=\frac{1}{2^N}\sum_{\vec{m}|m_k\in\{0,N+1\}}c_{\vec{m}}\bigotimes_{k\in V_Q}\hat{R}_{m_{k}}\bigotimes_{k\in V_c}\hat{R}_{m_{k}} +\frac{1}{2N}\sum_{j=1}^{N}(-1)^j\bigotimes_{k\in V_Q;m_k=j}\hat{R}_{m_k}\bigotimes_{k\in V_c;m_k=j}\hat{R}_{m_k}.\label{gn}
\end{eqnarray}
\\

\noindent(Step 2) All local measurement outcomes on the classical nodes are completely described by the pre-existing data: \{$\textbf{v}_{k}|{k\in V_{c}}\}$ [Eq.~(\ref{aai})]. Therefore, according Eq.~(\ref{gn}), the state fidelity function is represented as 
\begin{eqnarray}
F(N)=\frac{1}{2^N}\sum_{\vec{m}|m_k\in\{0,N+1\}}\!\!c_{\vec{m}}\!\left\langle\bigotimes_{k\in V_Q}\hat{R}_{m_{k}}\right\rangle\!\!\left\langle\prod_{k\in V_c}R_{m_{k}}\right\rangle +\frac{1}{2N}\!\sum_{j=1}^{N}(-1)^j\!\!\left\langle\bigotimes_{k\in V_Q;m_k=j}\hat{R}_{m_k}\right\rangle\!\!\left\langle\prod_{k\in V_c;m_k=j}R_{m_k}\right\rangle.\label{aaf}
\end{eqnarray}
 
\noindent(Step 3) With Eq.~(\ref{aaf}), Eq.~(\ref{aal}) becomes
\begin{equation}
\mathcal{F}_{n_{c}}(N)=\max_{\{\textbf{v}_{k}| k\in V_{c},|V_{c}|=n_c\}}E[\hat{F}_{n_c}(N)],\label{aap}
\end{equation} 
where $E[\hat{F}_{n_c}(N)]$ denotes the maximum eigenvalue of the operator:
\begin{eqnarray}
\hat{F}_{n_c}(N)=\frac{1}{2^N}\sum_{\vec{m}|m_k\in\{0,N+1\}}c_{\vec{m}}\bigotimes_{k\in V_Q}\hat{R}_{m_{k}}\left\langle\prod_{k\in V_c}R_{m_{k}}\right\rangle +\frac{1}{2N}\sum_{j=1}^{N}(-1)^j\bigotimes_{k\in V_Q;m_k=j}\hat{R}_{m_k}\left\langle\prod_{k\in V_c;m_k=j}R_{m_k}\right\rangle.\label{fnc}
\end{eqnarray}
This is the fidelity operator we use in Methods for showing false positives of EW-based entanglement detection in the main text. It can be rephrased further as
\begin{equation}
\mathcal{F}_{n_{c}}(N)=E[\hat{F}^{\max}_{n_c}(N)],\label{aap}
\end{equation} 
where $\hat{F}^{\max}_{n_c}(N)$ is the operator which can provide the maximum eigenvalue of $\hat{F}_{n_c}(N)$ under specific conditions of $V_c$ and $\{\textbf{v}_{k}| k\in V_{c}\}$. In the following, we give two concrete examples and then show general cases of $\mathcal{F}_{n_{c}}(N)$.\\ 

\subsubsection*{Example of $N=3$ and $n_c=1$}

Following (Step 1)-(Step 3) and $\textbf{v}_{3}=\textbf{v}_{3,1}$ [see Eq.~(\ref{v3})] we have
\begin{eqnarray}
\mathcal{F}_{1}(3)&=&E[ \frac{1}{8}( I \otimes I (1)+  I \otimes Z (1) +  Z \otimes Z (1)+  Z \otimes I (1))\nonumber\\
&&\ \ \ + \frac{1}{6}(-\hat{R}_{1} \otimes \hat{R}_{1} (1)+ \hat{R}_{2} \otimes \hat{R}_{2} (1)- \hat{R}_{3} \otimes\hat{R}_{3} (1))]\nonumber\\
&\simeq& 0.667.
\end{eqnarray} 

\subsubsection*{Example of $N=4$ and $n_c=2$}

With the same method used above, given $\textbf{v}_{3}=\textbf{v}_{3,12}$ and $\textbf{v}_{4}=\textbf{v}_{4,1}$ [see Eq.~(\ref{v4})] we have
\begin{eqnarray}
\mathcal{F}_{2}(4)&=&E[ \frac{1}{16}(I\otimes I(1)(1)+ I\otimes I(1)(1) + I\otimes Z(1)(1)+ I\otimes Z(1)(1)\nonumber\\
&&\ \ \ + Z\otimes I(1)(1)+ Z\otimes Z(1)(1)+ Z\otimes I(1)(1)
+ Z\otimes Z(1)(1))\nonumber\\
&&\ \ \ + \frac{1}{8}(-
\hat{R}_{1}\otimes\hat{R}_{1}(1)(1)+\hat{R}_{2}\otimes\hat{R}_{2}(-1)(1)-\hat{R}_{3}\otimes\hat{R}_{3}(-1)(1)+\hat{R}_{4}\otimes\hat{R}_{4}(1)(1))]\nonumber\\
&\simeq& 0.683.\label{f24}
\end{eqnarray}

\subsubsection*{General cases}

For more illustrations, the calculation results of state fidelity ${\mathcal{F}}_{n_c}(N)$ for $N=2,3,...,6$ and the corresponding $n_c$ can be seen in Supplementary Table~\ref{tab:1}.

\begin{table}[h]
\centering 
\begin{ruledtabular}
\begin{tabular}{cccccc}
$\mathcal{F}_{n_c}(N)$ & $N$ = 2 & $N$ = 3 & $N$ = 4 & $N$ = 5 & $N$ = 6\\
\hline
$n_c = 1$\  & 0.683& 0.667& 0.6613& 0.6589& 0.6576\\

$n_c = 2$\  & -& 0.667& 0.683& 0.6589& 0.667\\

$n_c = 3$\  & -& -& 0.6613& 0.6589& 0.683\\

$n_c = 4$\  & -& -& -& 0.6589& 0.667\\

$n_c = 5$\  & -& -& -& -& 0.6576
\end{tabular}
\end{ruledtabular}
%\justifying 
\caption{$\mathcal{F}_{n_c}(N)$ for $N=2,3,...,6$ and $1\leq n_c \leq N-1$.}
\label{tab:1}
\end{table}

As shown in Supplementary Table~\ref{tab:1}, we know that when the number of nodes $N$ is odd, the maximum state fidelity $\mathcal{F}_{n_{c}}(N)$ is identical regardless of the number of classical nodes $n_c$. Whereas, when the number of nodes $N$ is even, as long as the number of classical nodes $n_c$ is half the number of nodes, the maximum state fidelity of $\mathcal{F}_{n_{c}}(N)$ is 0.683 for $n_c=N/2$. The maximum state fidelity then symmetrically decreases with the increasing and decreasing number of classical nodes. For example, when $N=6$, we have ${\mathcal{F}}_{3}(6)\simeq 0.683>{\mathcal{F}}_{2}(6)={\mathcal{F}}_{4}(6)\simeq 0.667>{\mathcal{F}}_{1}(6)={\mathcal{F}}_{5}(6) \simeq 0.6576$. In addition to the above observations, from Supplementary Table~\ref{tab:1} we also notice that $\mathcal{F}_{1}(N)=\mathcal{F}_{N-1}(N)$ for arbitrary $N$. This can be confirmed for large $N$. See Supplementary Table~\ref{tab:2}.\\

\begin{table}[h!]
\begin{ruledtabular}
\begin{tabular}{ccccccc}

$\mathcal{F}_{n_c}(N)$ &$N$ = 7 & $N$ = 8 & $N$ = 9 & $N$ = 10 & $N$ = 11 & $N$ = 12 \\
\hline
${\mathcal{F}}_{1}(N)$ &0.6569  & 0.6564 & 0.6560 & 0.6558 & 0.6556 & 0.6555\\ 

${\mathcal{F}}_{N-1}(N)$ &0.6569  & 0.6564 & 0.6560 & 0.6558 & 0.6556 & 0.6555 
\end{tabular}
\end{ruledtabular}
\caption{$\mathcal{F}_{n_c}(N)$ for $n_c=1,N-1$ and $N=7,8,...,12$.}
\label{tab:2}
\end{table}

\subsection*{Deriving $\mathcal{F}_{c}(N)$ from $\mathcal{F}_{n_{c}}(N)$}\label{fcfnc}

For a given $N$, we consider all cases where the classical nodes exist in the created network and define the maximum value of the calculated maximum fidelities $\mathcal{F}_{n_{c}}(N)$ as the fidelity upper bound $\mathcal{F}_{c}(N)$.
The fidelity upper bound describes the minimum deviation between the $N$-node network with classical nodes and the ideal target state.

According to the regularity of $\mathcal{F}_{n_{c}}(N)$ in Supplementary Table~\ref{tab:1}, we have $\mathcal{F}_{c}(N) = \mathcal{F}_{n_{c}}(N)$ for odd $N$, and $\mathcal{F}_{c}(N) = \mathcal{F}_{N/2}(N)$ for even $N$. In what follows, we will describe how we obtain the analytical forms of $\mathcal{F}_{c}(N)$ for GHZ states from $\mathcal{F}_{n_{c}}(N)$, as given in Eq.~(6) in the main text.

\subsubsection*{Fidelity criteria for odd $N$}

The calculation of $\mathcal{F}_{c}(N)$ consists of the following three steps:\\

\noindent(Step 1) The matrix of the operator $\hat{F}_{n_c}(N)$ (\ref{fnc}) has a size of $2^{N-n_{c}}\times2^{N-n_{c}}$ and varies with $N$. Therefore, when $n_{c}=N-1$, its size becomes the minimal $2\times2$. The operator $\hat{F}_{N-1}(N)$ is used for further analysis.

\noindent(Step 2) As $\left\langle\prod_{k\in V_c}R_{m_{k}}\right\rangle=c_{\vec{m}}$ and $\left\langle\prod_{k\in V_c;m_k=j}R_{m_k}\right\rangle=(-1)^j$, the following operator can provide the maximum eigenvalue of $\hat{F}_{N-1}(N)$:
\begin{eqnarray}
\hat{F}^{\max}_{N-1}(N)=\frac{1}{2^N}\sum_{\vec{m}|m_k\in\{0,N+1\}}c^2_{\vec{m}}\hat{R}_{m_{k'}}+\frac{1}{2N}\sum_{j=1}^{N}\hat{R}_{m_{k'}=j}.
\end{eqnarray}
Alternatively, it has the matrix form:
\begin{equation}
\hat{F}^{\max}_{N-1}(N)=
\left(\begin{array}{cc}
\frac{1}{2} & 
\sum^{N}_{j=1}\frac{\cos({\frac{j\pi}{N}})
-i{\rm sin}({\frac{j\pi}{N})}}{2N} \\ 
\sum^{N}_{j=1}\frac{{\rm cos}({\frac{j\pi}{N}})+i{\rm sin}({\frac{j\pi}{N})}}{2N} & 0\end{array}\right).\label{aas}
\end{equation}
The off-diagonal terms are derived from the basic definitions of $\hat{R}_{m_{k'}=j}$ for $j=1,2,...,N$. For the diagonal matrix elements, there are $2^{N-2}$ permutations of $I+Z=2\left|0\right\rangle\!\!\left\langle 0\right|$ with the factor $1/2^N$, we therefore have $2^{N-2}\times 2\times 2^{-N}=1/2$, which is independent of $N$.

\noindent(Step 3) The maximum eigenvalue of $\hat{F}^{\max}_{N-1}(N)$ for $\mathcal{F}_{N-1}(N)$ is $(1/2+\sqrt{1/4+\frac{{\rm csc}^2(\frac{\pi}{2N})}{N^2}})/2$. Therefore, we have $\mathcal{F}_{c}(N)=\mathcal{F}_{N-1}(N)$ for odd $N$, as shown in Eq.~(6) in the main text.

\subsubsection*{Fidelity criteria for even $N$}\label{even}

Lu $et$ $al$. \cite{Lu20} prove that, as the Pauli matrices used as the observables, the maximum classical state fidelity occurs as the existence of one classical node in an $N$-node network and is ${\mathcal F}_{P,1}(N)\simeq 0.683$. Suppose that the maximum eigenvalue of $\hat{F}^{\max}_{N/2}(N)$ is ${\mathcal F}_{N/2}(N)$, and the maximum eigenvalue of $\hat{F}^{\max}_{P,1}(N)$ is ${\mathcal F}_{P,1}(N)$.
To confirm the regularity of ${\mathcal F}_{N/2}(N)\simeq 0.683$ for arbitrary even $N$, we prove that $\hat{F}^{\max}_{N/2}(N)=\hat{F}^{\max}_{P,1}(N/2+1)$, which shows that the maximum eigenvalue of $\hat{F}^{\max}_{N/2}(N)$ for arbitrary even $N$ is 0.683. The analysis process consists of the following two steps:\\

\noindent(Step~1) The unrepeatable operators related to $I$ and $Z$ in the diagonal terms in $\hat{F}^{\max}_{N/2}(N)$ and $\hat{F}^{\max}_{P,1}(N/2+1)$ are the same. According to Eq.~(\ref{aaf}), there are $2^{N-1}$ permutations of operators related to $I$ and $Z$ in the ideal projector of GHZ states. For example, see Eq.~(\ref{gm4}). However, since the measurement outcomes on the classical nodes are described by a pre-existing fixed set \{$\textbf{v}_{k}|{k\in V_{c}}\}$ [Eq. (\ref{aai})], the terms with the same forms appear $2^{\frac{N}{2}-1}$ times in the operator $\hat{F}^{\max}_{N/2}(N)$. Therefore, the number of unrepeatable operators related to $I$ and $Z$ for the diagonal terms in $\hat{F}^{\max}_{N/2}(N)$ is $2^{N-1}/2^{\frac{N}{2}-1}=2^{\frac{N}{2}}$. The $\hat{F}^{\max}_{P,1}(N/2+1)$ also has $2^{\frac{N}{2}}$ unrepeatable terms involving $I$ and $Z$. Each unrepeatable term has the amplitude: $2^{-N}$ [the normalization factor in Eq. (\ref{aaf})] $\times2^{\frac{N}{2}-1}=2^{-\frac{N}{2}-1}$.

For example, there are 8 terms involving $I$ and $Z$ in $\hat{F}^{\max}_{2}(4)$ [see Eq.~(\ref{f24})]. The existing unrepeatable four terms are: $II$, $IZ$, $ZI$, and $ZZ$, where each one appears twice in $\hat{F}^{\max}_{2}(4)$. These four terms match the unrepeatable terms involving $I$ and $Z$ in $\hat{F}^{\max}_{P,1}(3)$. The amplitude of these terms is $2^{-\frac{4}{2}-1}=1/8$. For $N=6$, we have the 8 unrepeatable terms $III$, $IIZ$, $IZI$, $ZII$, $ZZI$, $ZIZ$, $IZZ$, $ZZZ$, where each one appears 4 times in $\hat{F}^{\max}_{3}(6)$ with total 32 terms and has the amplitude $2^{-\frac{6}{2}-1}=1/16$, and these 8 terms and their amplitudes match the terms involving $I$ and $Z$ in $\hat{F}^{\max}_{P,1}(4)$.

\noindent(Step 2) We analyze the off-diagonal terms of $\hat{F}^{\max}_{N/2}(N)$ and obtain all $X$ and $Y$ terms in $\hat{F}^{\max}_{P,1}(N/2+1)$ by expanding the observable $\hat{R}_{m_{k}}$ for $m_k=1,2,...,N$ in $\hat{F}^{\max}_{N/2}(N)$. The off-diagonal terms of  $\hat{F}^{\max}_{N/2}(N)$ [see Eq.~(\ref{fnc})] is $\frac{1}{2N}\sum_{j=1}^{N}(-1)^j\bigotimes_{k\in V_Q;m_k=j}\hat{R}_{m_k}\left\langle\prod_{k\in V_c;m_k=j}R_{m_k}\right\rangle$. Since these observables $\hat{R}_{m_{k}}$ in $\hat{F}^{\max}_{N/2}(N)$ are composed of operators related to $X$ and $Y$ only, we find that these terms are the same as those of $\hat{F}^{\max}_{P,1}(N/2+1)$ under specific measurement outcomes $\left\langle\prod_{k\in V_c;m_k=j}R_{m_k}\right\rangle$ of the classical nodes.

As expanding the observable $\hat{R}_{m_{k}}$ and represent $\hat{F}^{\max}_{N/2}(N)$ in $X$ and $Y$, one can find the coefficients of the operators involving $X$ and $Y$ in $\hat{F}^{\max}_{N/2}(N)$ and $\hat{F}^{\max}_{P,1}(N/2+1)$ are identical. Let us take the terms of $X^{\otimes\frac{N}{2}}$ and $Y^{\otimes\frac{N}{2}}$ in $\hat{F}^{\max}_{N/2}(N)$ for example. According to the definitions of $\hat{R}_{m_{k}}$ and $\hat{F}_{N/2}(N)$, the coefficients of $X^{\otimes\frac{N}{2}}$ are
\begin{equation}
\frac{1}{2N}\sum_{j=1}^{N}(-1)^j\prod_{k\in V_Q;m_k=j}{\rm cos} (\frac{m_{k}\pi}{N})\left\langle\prod_{k\in V_c;m_k=j}R_{m_k}\right\rangle=\pm \frac{1}{2^{\frac{N}{2}+1}},\label{aax}
\end{equation}
and the coefficients of $Y^{\otimes\frac{N}{2}}$ are
\begin{equation}
\frac{1}{2N}\sum_{j=1}^{N}(-1)^j\prod_{k\in V_Q;m_k=j}{\rm sin} (\frac{m_{k}\pi}{N})\left\langle\prod_{k\in V_c;m_k=j}R_{m_k}\right\rangle=\pm \frac{1}{2^{\frac{N}{2}+1}}.\label{aay}
\end{equation}
The terms $X^{\otimes\frac{N}{2}}$ and $Y^{\otimes\frac{N}{2}}$ in $\hat{F}^{\max}_{P,1}(N/2+1)$ have the same coefficients. Equations (\ref{aax}) and (\ref{aay}) can be satisfied at the same time only for even $N$ and $n_{c} = N/2$ such that the maximum eigenvalue of $\hat{F}^{\max}_{n_c}(N)$ is 0.683 (see Supplementary Table~\ref{tab:1}). Since the maximum eigenvalue of $\hat{F}^{\max}_{P,1}(N/2+1)$ is determined to be $0.683$ by Lu \textit{et al.}~\cite{Lu20}, the maximum eigenvalue obtained from $\hat{F}^{\max}_{N/2}(N)$ is determined to be 0.683 as well. Moreover, according to the regularity of the maximum state fidelity as shown in Table~\ref{tab:1}, when the number of nodes $N$ is even and the number of classical nodes is $n_{c} = N/2$, the maximum classical state fidelity for the fidelity criteria [Eqs.~(5)] are
\begin{eqnarray}
\mathcal{F}_{c}(N)\simeq 0.683,
\label{aaz}
\end{eqnarray}
for arbitrary even $N$.

\subsubsection*{Example of the case of $N=4$}

By considering (Step~1) and (Step~2) in the above section~\ref{even}, we have $\hat{F}^{\max}_{2}(4)=\hat{F}^{\max}_{P,1}(3)$. We assume that the 3rd and 4th qubits decay into the classical nodes. According to the above decomposition of $\hat{F}^{\max}_{2}(4)$ (\ref{f24}), the coefficient of $X^{\otimes\frac{4}{2}}$ is
\begin{eqnarray}
\frac{1}{2\times 4}\sum_{j=1}^{4}(-1)^j\prod_{k\in V_Q;m_k=j}{\rm cos} (\frac{m_{k}\pi}{N})\left\langle\prod_{k\in V_c;m_k=j}R_{m_k}\right\rangle&=&\frac{1}{8} [(-1){\rm cos} (\frac{1 \pi}{4})^{{2}}
\left\langle\prod_{k\in V_{c}}{R}_{1}\right\rangle+(1){\rm cos} (\frac{2\pi}{4})^{{2}}
\left\langle\prod_{k\in V_{c}}{R}_{2}\right\rangle\nonumber\\
&&\ \ +(-1){\rm cos} (\frac{3\pi}{4})^{{2}}
\left\langle\prod_{k\in V_{c}}{R}_{3}\right\rangle+(1){\rm cos} (\frac{4\pi}{4})^{{2}} \left\langle\prod_{k\in V_{c}}{R}_{4}\right\rangle]\nonumber.
\end{eqnarray}
${\left\langle\prod_{k\in V_{c}}{R}_{m_{k}}\right\rangle}_{\vec{m}}$ are described by sets $\textbf{v}_{3,1}$ and $\textbf{v}_{4,12}$, that is, 
$\left\langle\prod_{k\in V_{c}}{R}_{1}\right\rangle
=(1)\times(1)=1$,
$\left\langle\prod_{k\in V_{c}}{R}_{2}\right\rangle
=(1)\times(-1)=-1$,
$\left\langle\prod_{k\in V_{c}}{R}_{3}\right\rangle
=(1)\times(-1)=-1$,
$\left\langle\prod_{k\in V_{c}}{R}_{4}\right\rangle
=(1)\times(1)=1$. 
The coefficient of $X^{\otimes\frac{4}{2}}$ therefore is
\begin{eqnarray}
&&\frac{1}{8}[(-1){\rm cos} (\frac {1\pi}{4})^{{2}}(1)+(1){\rm cos} (\frac{2\pi}{4})^{{2}}(-1)+(-1){\rm cos} (\frac{3\pi}{4})^{{2}}(-1)+(1){\rm cos} (\frac{4\pi}{4})^{{2}} (1)]= \frac{1}{8}\nonumber,
\end{eqnarray}
which is equal to the coefficient of the $X^{\otimes 2}$ term in $\hat{F}^{\max}_{P,1}(3)$.

Similarly, the coefficient of $Y^{\otimes\frac{4}{2}}$ term is
\begin{eqnarray}
\frac{1}{2\times 4}\sum_{j=1}^{4}(-1)^j\prod_{k\in V_Q;m_k=j}{\rm sin} (\frac{m_{k}\pi}{N})\left\langle\prod_{k\in V_c;m_k=j}R_{m_k}\right\rangle&=&\frac{1}{8} [(-1){\rm sin} (\frac{1 \pi}{4})^{{2}}
\left\langle\prod_{k\in V_{c}}{R}_{1}\right\rangle+(1){\rm sin} (\frac{2\pi}{4})^{{2}}
\left\langle\prod_{k\in V_{c}}{R}_{2}\right\rangle\nonumber\\
&&\ \ +(-1){\rm sin} (\frac{3\pi}{4})^{{2}}
\left\langle\prod_{k\in V_{c}}{R}_{3}\right\rangle+(1){\rm sin} (\frac{4\pi}{4})^{{2}} \left\langle\prod_{k\in V_{c}}{R}_{4}\right\rangle]\nonumber.
\end{eqnarray}
where $\left\langle\prod_{k\in V_{c}}{R}_{1}\right\rangle$, $\left\langle\prod_{k\in V_{c}}{R}_{2}\right\rangle$, $\left\langle\prod_{k\in V_{c}}{R}_{3}\right\rangle$,
and $\left\langle\prod_{k\in V_{c}}{R}_{4}\right\rangle$ are set the same as those for $X^{\otimes\frac{4}{2}}$ term. Then we have the coefficient of $Y^{\otimes\frac{4}{2}}$:
\begin{eqnarray}
\frac{1}{8}[(-1){\rm sin} (\frac{1\pi}{4})^{2}
(1)+(1){\rm sin} (\frac{2\pi}{4})^{2}
(-1)+(-1){\rm sin} (\frac{3\pi}{4})^{2}
(-1)+(1){\rm sin} (\frac{4\pi}{4})^{2} (1)]=-\frac{1}{8},\nonumber
\end{eqnarray}
which is equal to the coefficient of the $Y^{\otimes 2}$ term in $\hat{F}^{\max}_{P,1}(3)$.

In addition to the diagonal terms, $X^{\otimes2}$ and $Y^{\otimes2}$, the same method is used to analyze the terms of $X\otimes Y$ and $Y\otimes X$. Thus $\hat{F}^{\max}_{2}(4)$ represented in the Pauli matrices is
\begin{eqnarray}
\hat{F}^{\max}_{2}(4) &=& 
\frac{1}{8}(I \otimes I(1)
+ I\otimes Z(1)+ Z\otimes Z(1)
+ Z\otimes I(1)\nonumber\\
&&+ X\otimes X(1)
- Y\otimes Y(1)- Y\otimes X(1)
- X\otimes Y(1)).\nonumber
\end{eqnarray}
For $\hat{F}^{\max}_{P,1}(3)$, if we assume that the third node is classical and described by a pre-existing measurement data by $\textbf{v}_{3,1}$ (\ref{v3}), we obtain the same result as $\hat{F}^{\max}_{2}(4)$, and Eq.~(\ref{G3_pauli}) becomes:
\begin{eqnarray}
\hat{F}^{\max}_{P,1}(3) &=& \frac{1}{8}(I \otimes I(1)+ I\otimes Z(1) + Z\otimes Z(1)+ Z\otimes I(1)\nonumber\\
&&+ X\otimes X(1)- Y\otimes Y(1)- Y\otimes X(1)- X\otimes Y(1)).\nonumber
\end{eqnarray}
For $\hat{F}^{\max}_{2}(4)=\hat{F}^{\max}_{P,1}(3)$, the maximum eigenvalue of $\hat{F}^{\max}_{P,1}(3)$ implies that $\mathcal{F}_{{2}}(4)\simeq 0.683$ [Eq.~(\ref{aaz})].\\

\section*{SUPPLEMENTARY NOTE 4: Noise tolerance of steering criteria} 

To compare the steering criteria [Eq.~(4)] with the ones based on $2^{N-1}+1$ measurement settings for Pauli observables \cite{Lu20}, we examine their robustness against noise, i.e., noise tolerance. This is evaluated by considering the maximum intensity of white noise added to $\left|{G_N}\right\rangle$ such that genuine multipartite EPR steerability of the resulting mixed state start to become undetectable by the criterion:
\begin{equation}
F(N,p_{\text{noise}})=\text{tr}(\left|{G_N}\right\rangle\!\!\left\langle {G_N}\right| \rho_{N}(p_{\text{noise}}))=\mathcal{F}_{c}(N),
\end{equation}
where
\begin{equation}
\rho_{N}(p_{\text{noise}})= p_{\text{noise}}\frac {I^{\otimes N}}{2^{N}}+
(1-p_{\text{noise}})\left|{G_N}\right\rangle\!\!\left\langle {G_N}\right|,
\label{eea}
\end{equation}
and $p_{\text{noise}}$ is called the threshold intensity of the white noise. Therefore a steering criterion possesses higher noise tolerance than another criterion if it has a larger threshold noise intensity $p_{\text{noise}}$.

The noise tolerance of the steering criteria with $N+1$ measurement settings is shown in Supplementary Figure~\ref{fig_s_a} for $N=2,3,...,14$. Compared to the noise tolerance of the existing criteria \cite{Lu20}, the present efficient steering criteria even have better noise tolerance for the cases of odd $N$. The reason is that $\mathcal{F}_c$ [Eq.~(5) in the main text] are lower than the classical fidelity upper bounds (0.683) in their criteria. For the cases of even $N$, since the classical fidelity upper bounds are identical in both types of criteria, they have the same noise tolerance.\\

\begin{figure}[t!]
\centering
\includegraphics[width=0.45\columnwidth]{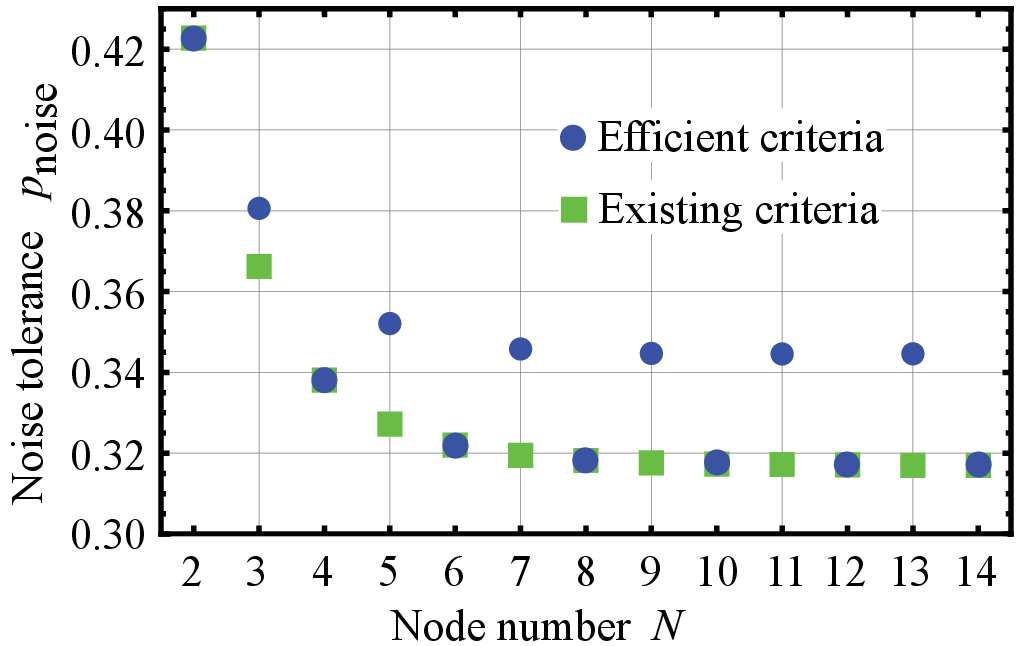}
\caption{The noise tolerance of the efficient steering criteria (blue) and the existing steering criteria (green) \cite{Lu20} for $N=2,3,...,14$.}
\label{fig_s_a}
\end{figure}

\section*{SUPPLEMENTARY NOTE 5: Comparison of the present work and the previous work}
In this section, we aim to discuss the difference between the present work and the previous work, Lu \textit{et al.} \cite{Lu20}. For both works, the underlying concept and method to derive the maximum fidelity of networks with classical nodes depend on two items: (1) the observables used for the state decomposition of the target state, and (2) the calculation of the maximum fidelity. In particular, the maximum fidelity a quantum-classical hybrid system can attain depends on what observables the state projector is decomposed into. The decomposition observables utilized in the work of Lu \textit{et al.} \cite{Lu20} are rather different from the ones in our state decomposition. Their observables for a quantum node are the Pauli matrices. They form an orthonormal set of matrices with respect to the Hilbert--Schmidt inner product. Therefore, such or=thonormal characteristics make the state projector decomposition like a usual quantum state tomography, employing an orthonormal set of matrices. In contrast, the observables used in the present work for the state composition do not possess this orthogonality property. When using the observables with the orthogonality property to decompose the target state and obtain the corresponding fidelity function, such characteristics can cause the maximum fidelity of the quantum-classical hybrid networks to decrease with the number of classical nodes monotonically. This means a one-to-one correspondence exists between the number of classical nodes and the relevant maximum fidelity values, providing a set of graduations to indicate the degree of network imperfection. Without the orthonormal observables, the maximum fidelities do not possess the strictly decreasing feature for an arbitrary node number. However, the cost of having a set of graduations to indicate the degree of network imperfection requires measurement settings exponentially increasing with the network node number. On the contrary, ours only requires a linearly increasing number of measurement settings.

Since the observables utilized in the state decomposition in our formalism are not orthogonal, the maximal fidelity thresholds do not monotonically decrease with the classical node number for a given total node number. Explicitly, the maximum fidelity remains and does not change with the classical node number under a given odd total node number. Maximum fidelity changes with the classical node number for a given even total node number, but the fidelity value does not decrease monotonically. Moreover, the fidelity value maximized over all classical node numbers does not change with the even total node number, whereas this maximum fidelity monotonically decreases with the odd total node number. See Supplementary Table~\ref{tab:1} for concrete examples. Although we understand the origin of the difference between our results and the existing ones is the orthogonality property of observables for state decomposition, we are still unclear as to why the behavior of the even total node number cases differs from the odd total node number cases observed in our present study. This unclear but important point will be one of the subjects in our future study.

\section*{SUPPLEMENTARY NOTE 6: Laser system and photon source for multi-photon GHZ states}

\begin{figure}[t]
\centering
\includegraphics[width=0.6\columnwidth]{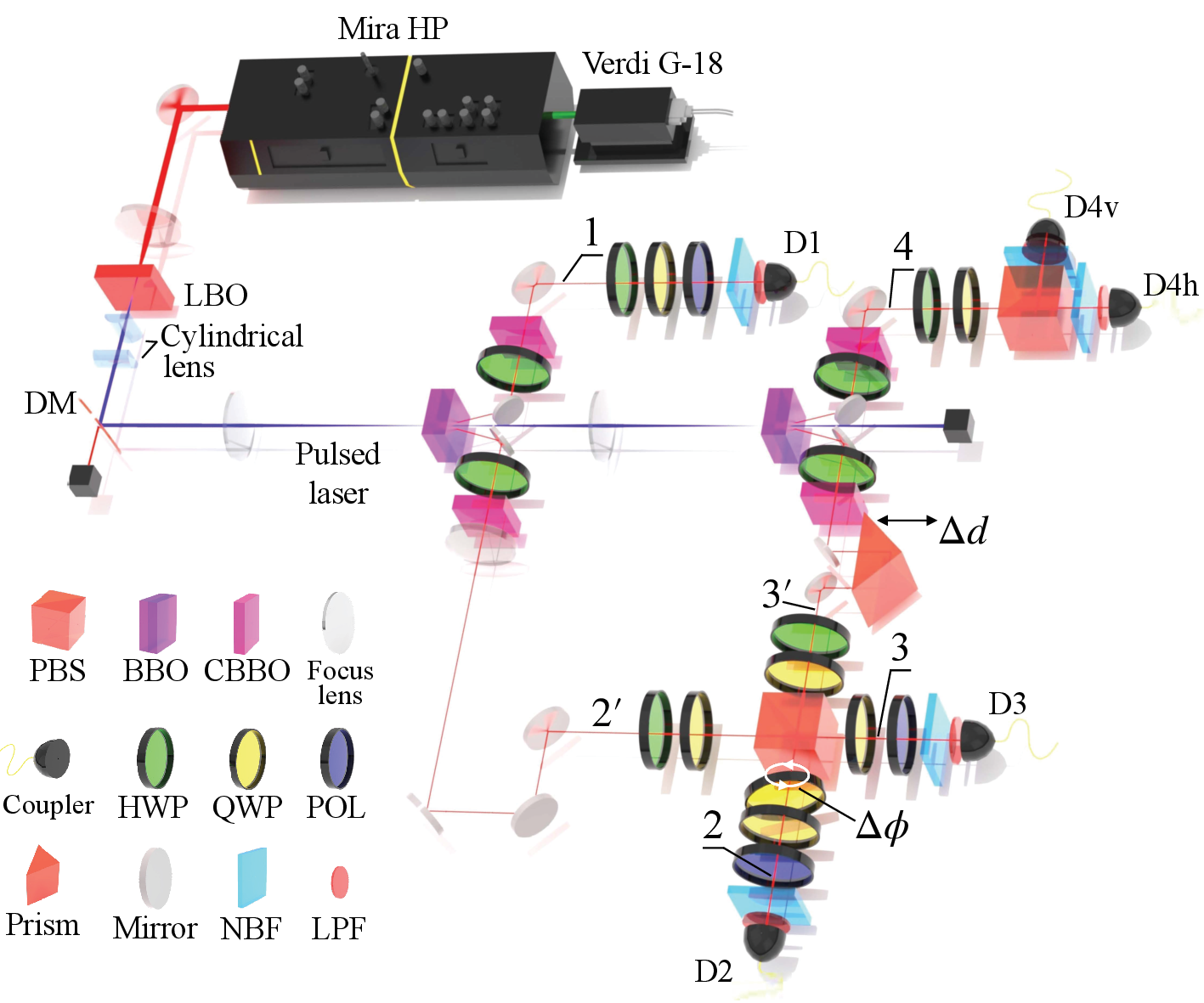}
\caption{A schematic diagram of the experimental set-up. See also Fig.~2 in the main text.}
\label{supsetup}
\end{figure}

Supplementary Figure~\ref{supsetup} shows a schematic of our laser system and multi-photon experimental set-up. First, we used a diode-pumped solid-state CW laser Verdi-G18 to pump a high-power ultrafast Ti:sapphire oscillator Mira HP. An ultrafast infrared pulsed laser beam was generated with an output power of about 2.5 W when Verdi-G18 operated at 13.5 W. The ultrafast infrared pulsed laser was focused on the up-converted crystal $\rm{LiB_{3}O_{5}}$ (LBO) to generate ultraviolet pulses via up-conversion. The ultraviolet laser beam was shaped to be circularized by the horizontal and vertical cylindrical lens. Then seven dichroic mirrors (DM) filtered out the infrared light and extracted the ultraviolet light. The resulting ultraviolet laser beam has a center wavelength of 390 nm, a pulse duration of 144 fs, a repetition rate of 76 MHz, and an average power of about 1.0 W.

Pairs of two-photon entanglement were generated by consecutively pumping a pulsed laser into two BBO crystals (2 mm). The entangled photon pairs passed through a birefringence compensator consisting of an HWP set at $45^{\circ}$ and BBO crystals with 1 mm thickness, which were used to eliminate the spatial and longitudinal walk-off effects of photons in modes $1$ and $2'$, modes $3'$ and 4, respectively.  The pairs of two-photon entanglement were prepared according to the target state of $\left|\Phi^{+}\right\rangle$ using HWPs.
The state fidelity is defined as $F_{s}=\rm{tr}(\rho_{{\rm expt},\Phi^+}\left|\Phi^+\right\rangle\!\!\left\langle \Phi^+\right|)$, where $\rho_{{\rm expt},\Phi^+}$ is the density operator of the created photon pairs. Since the state $\left|\Phi^{+}\right\rangle\!\!\left\langle \Phi^{+}\right|$ can be decomposed in terms of Pauli matrices as $\left|\Phi^{+}\right\rangle\!\!\left\langle \Phi^{+}\right|=(I\otimes I+X\otimes X-Y\otimes Y+Z\otimes Z)/4$, the state fidelity of $\left|\Phi^{+}\right\rangle$ can be written as $F_{s,\phi^{+}}=(1+\langle{X}{X}\rangle-\langle{Y}{Y}\rangle+\langle{Z}{Z}\rangle)/4$. Then we measured the state fidelity of $\left|\Phi^{+}\right\rangle$ by the three local measurements of $\langle{X}{X}\rangle$, $\langle{Y}{Y}\rangle$, and $\langle{Z}{Z}\rangle$. We obtained the above expectation values via measuring the coincidence counts of the eigenstates of the above measurements. We observe $\sim \rm{11.8\times 10^{4}}$ photon pairs in mode 1-$2'$ per second from the first BBO and $\sim \rm{8.0\times 10^{4}}$ photon pairs in mode $3'$-4 per second from the second BBO. The state fidelity of the first pairs is ${(\rm 92.23} \pm {\rm 0.09}) \%$, and the second pairs have the fidelity of $({\rm 94.98} \pm {\rm 0.08})\%$.\\

\section*{SUPPLEMENTARY NOTE 7: HOM-type interference}

To realize the Hong-Ou-Mandel (HOM)-type interference \cite{HOM87,Pan01,Pan12}, it is necessary to have a good spatial and temporal overlap for the two-photon indistinguishability.
For the spatial overlap, as shown in Fig.~2 in the main text and Supplementary Figure~\ref{supsetup}, first, we introduced one photon of each entangled pair, modes $2'$ and $3'$, into the $\rm{PBS}$ of modes $2$ and $3$, then spectrally filtered by the narrow-band filters (NBFs) and monitored by fiber-coupled single-photon detectors (D2 and D3) which are consisted of the adjustable fiber couplers to collect photons into the single-photon counting modules.
Since the path length of the down-converted photon $2'$ and the down-converted photon $3'$ are different, it is impossible to detect these two photons through the same focal length of the fiber couplers. We set a 10 cm focus lens in the path of photon $2'$ to refocus the down-converted photon to make it the same as the path of photon $3'$ to make it possible to be detected by the detectors, D2 and D3.

\begin{figure}[t]
\centering
\includegraphics[width=0.6\columnwidth]{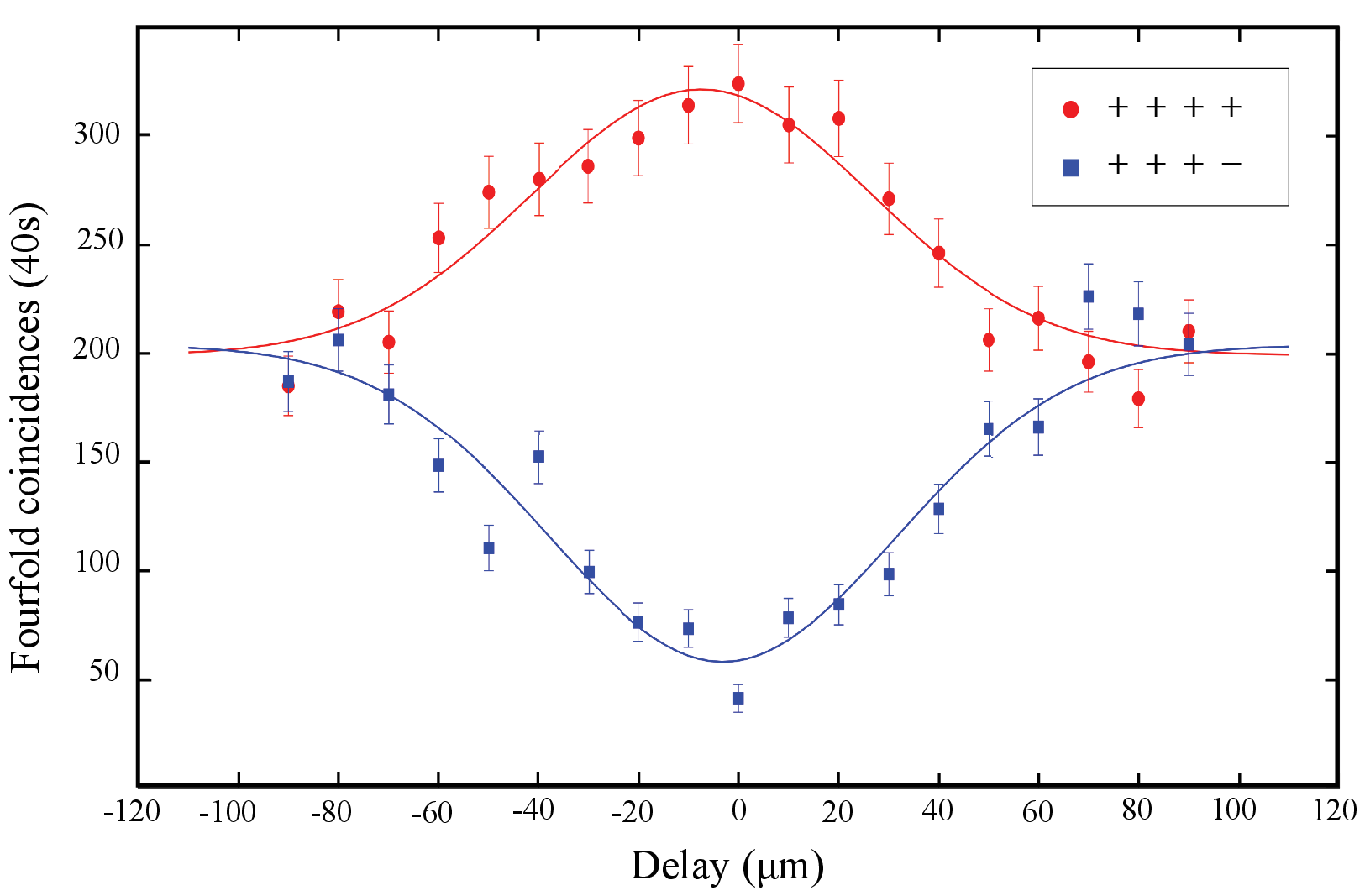}
\caption{Hong-Ou-Mandel-type (HOM-type) two-photon interference.}
\label{HOMdip}
\end{figure} 

To realize the temporal overlap, we adjust the path length of photon $2'$ via the delay prism to match the path length of photon $3'$.
We observed the HOM-type interference fringe in the diagonal [$\left|+\right\rangle=(\left|H\right\rangle+\left|V\right\rangle)/\sqrt{2}$/$\left|-\right\rangle=(\left|H\right\rangle-\left|V\right\rangle)/\sqrt{2}$] basis behind the PBS. With setting polarizers behind the PBS at $\rm{+45^\circ / -45^\circ}$, we measured the $\left|+\right\rangle\left|+\right\rangle\left|+\right\rangle\left|+\right\rangle$ and $\left|+\right\rangle\left|+\right\rangle\left|+\right\rangle\left|-\right\rangle$ coincidence counts between detectors D1, D2, D3, and D4 (D4h and D4v) for the HOM-type fringe. The HOM-type interference fringe is shown in Supplementary Figure~\ref{HOMdip}. The interferometer is sensitive to only the order of magnitude of the length change to the detected photon coherence length $\rm{\sim 100~\mu m}$. Here, the coherence length is estimated by the standard deviation of the coincidence counts for the length change of the delay.

\section*{SUPPLEMENTARY NOTE 8: Experimental three-photon GHZ state and state fidelities measured with Pauli observables}
The three-photon GHZ state $\left|G_{3}\right\rangle$ was experimentally created as follows. First, the first pair of entangled states $\left|\Phi^{+}\right\rangle$ was measured on photon 1 in the state $V$, which acts as a trigger for preparing $\left|G_{3}\right\rangle$. See Supplementary Figure~\ref{supsetup}. Then, to prepare photon $2'$ in the state $\left|+\right\rangle$, it passed through a HWP with an optical axis setting at $22.5^{\circ}$ with respect to the vertical axis, and a QWP with optical axis setting at $45^{\circ}$ with respect to the vertical axis. Finally, photon $2'$ interfered with photon $3'$ of the second pair of entangled state $\left|\Phi^{+}\right\rangle$ to generate a state $\rho_{{\rm expt},G3}$ close to the three-photon GHZ state $\left|G_{3}\right\rangle$. With Eq.~(\ref{G3_pauli}), we measured the state fidelity of by experimentally determining the five expectation values: $\langle R_1R_1R_1 \rangle $, $\langle R_1R_2R_2 \rangle $, $\langle R_3R_3R_3\rangle$, $\langle R_2R_1R_2\rangle$, and $\langle R_2R_2R_1\rangle$, were photons in mode 4 were measured by HWP and QWP, and photons in modes 2 and 3 were measured with QWP and polarizer (POL). Here $R_1$, $R_2$, and $R_3$ correspond to the Pauli observables $\hat{R}_1=X$, $\hat{R}_2=Y$, and $\hat{R}_3=Z$, respectively. See the experimental expectation values of observables in Supplementary Figure~\ref{expected_values_3}. The resulting fidelity of $\rho_{{\rm expt},G3}$ is $(77.93\pm 0.62)\%$. The errors are calculated by Poissonian counting statistics. With the fidelity criterion by Lu \textit{et al.} \cite{Lu20}, our experimental four-photon and three-photon states possess genuine multipartite EPR steerability, which are better than the classical fidelity upper bound of 0.683. Moreover, this result is comparable to the experimental fidelity of $(78.00\pm0.60)\%$ measured with 4 local measurement settings (\ref{gm3}) [Fig.~3(a) in the main text].

\begin{figure}[t]
\centering
\includegraphics[width=0.6\columnwidth]{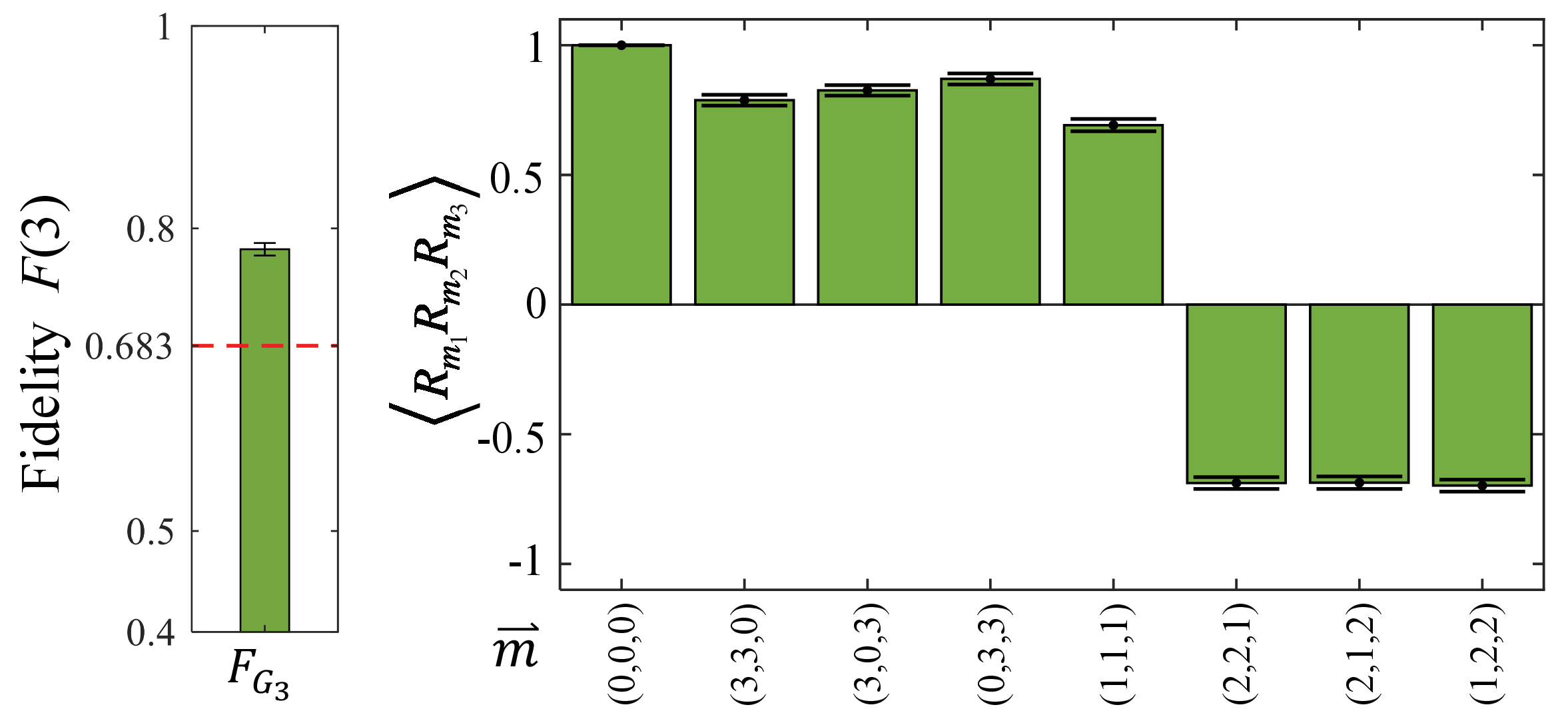}
\caption{Expectation values of Pauli observables for $\rho_{{\rm expt},G3}$. Here we have defined that $\hat{R}_0=I,\hat{R}_1=X,\hat{R}_2=Y,\hat{R}_3=Z$.}
\label{expected_values_3}
\end{figure} 

\begin{figure}[t]
\centering
\includegraphics[width=0.8\columnwidth]{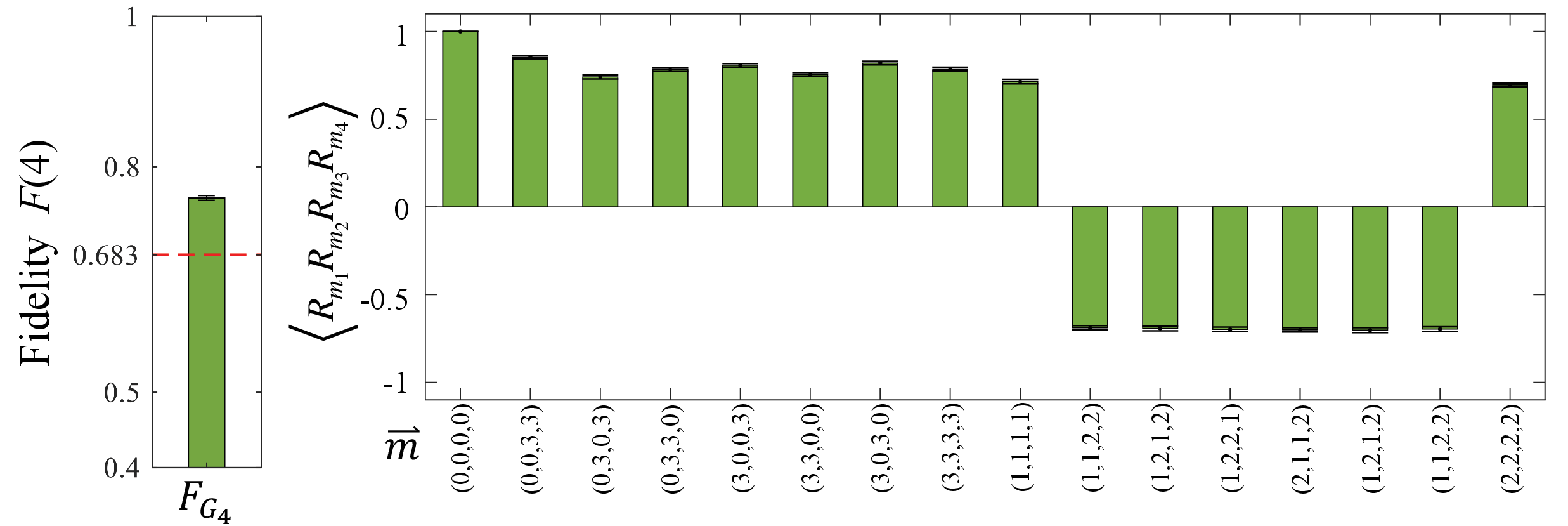}
\caption{Expectation values of Pauli observables for $\rho_{{\rm expt},G4}$.}
\label{expected_values_4}
\end{figure} 

Equation~(\ref{G4_pauli}) implies that the state fidelity can be experimentally determined by carrying out the nine local measurements for the following expectation values: $\langle R_1R_1R_1R_1 \rangle $, $\langle R_2R_2R_2R_2 \rangle $, $\langle R_1R_1R_2R_2 \rangle $, $\langle R_1R_2R_2R_1 \rangle $, $\langle R_1R_2R_1R_2 \rangle$, $\langle R_2R_2R_1R_1 \rangle $, $\langle R_2R_1R_2R_1 \rangle $, and $\langle R_2R_1R_1R_2 \rangle $. In the experiment, photons 1 and 4 were measured by using HWP and QWP, and photons 2 and 3 were measured by using QWP and POL. We observed the four-fold coincidence counts for all the above measurements in a 4 ns coincidence window. We obtained $\sim 3023$ four-photon events per 40 seconds. The resulting experimental four-photon state fidelity for the experimental state $\rho_{{\rm expt},G4}$ is $(75.84\pm 0.31)\%$. According to the derived classical upper bound by Ref.~\cite{Lu20} of 0.683, $\rho_{{\rm expt},G4}$ possesses genuine four-node steerability close to $\left|G_4\right\rangle$. The measured expectation values of the observables are shown in Supplementary Figure~\ref{expected_values_4}. This result is similar to the state fidelity of $(75.79\pm 0.37)\%$ by using $5$ measurement settings, Eq.~(\ref{gm4}) [see also Fig.~3(b) in the main text].\\

\bibliography{SDMEQN}